\documentclass[%
 reprint,
superscriptaddress,
 amsmath,amssymb,
 aps, prl
]{revtex4-1}

\usepackage{graphicx}% Include figure files
\usepackage{dcolumn}% Align table columns on decimal point
\usepackage{bm}% bold math
\newcommand{\ii}{\mathrm{i}}
\newcommand{\tens}[1]{\overset{\text{\tiny$\leftrightarrow$}}{\mathbf{#1}}}

\renewcommand{\P}{{\rm P}}
\renewcommand{\S}{{\rm S}}
\newcommand{\I}{{\rm I}}
\usepackage{physics}
\usepackage{lipsum}
\usepackage{xcolor}
\usepackage[colorlinks=true,bookmarks=false,allcolors=blue]{hyperref}
\hypersetup{pdfstartview={FitH},pdfpagemode={UseNone}, breaklinks=true,            colorlinks,allcolors=blue, bookmarksopen=true, pdfnewwindow=true}
\usepackage[all]{hypcap}
\allowdisplaybreaks

\begin{document}

\preprint{APS/123-QED}

\title{Subdiffraction Quantum Imaging with Undetected Photons
%\\or\\ Subdiffraction Dark-Field Microscopy with Undetected Photons
}% Force line breaks with \\

\author{Elkin A. Santos}
\email{elkin.santos@uni-jena.de}
\affiliation{Institute of Applied Physics, Abbe Center of Photonics, Friedrich Schiller University Jena, Albert-Einstein-Str. 15, 07745 Jena,
Germany
}
\author{Thomas Pertsch}
\affiliation{Institute of Applied Physics, Abbe Center of Photonics, Friedrich Schiller University Jena, Albert-Einstein-Str. 15, 07745 Jena,
Germany
}
\affiliation{Fraunhofer Institute for Applied Optics and Precision Engineering IOF, Albert-Einstein-Str. 7, 07745 Jena, Germany}
\author{Frank Setzpfandt}
\affiliation{% 
Institute of Applied Physics, Abbe Center of Photonics, Friedrich Schiller University Jena, Albert-Einstein-Str. 15, 07745 Jena,
Germany
}%
\affiliation{Fraunhofer Institute for Applied Optics and Precision Engineering IOF, Albert-Einstein-Str. 7, 07745 Jena, Germany}
\author{Sina Saravi}%
\affiliation{Institute of Applied Physics, Abbe Center of Photonics, Friedrich Schiller University Jena, Albert-Einstein-Str. 15, 07745 Jena,
Germany
}


\date{\today}% It is always \today, today,

\begin{abstract}
We propose a nonlinear imaging scheme with undetected photons that overcomes the diffraction limit by transferring near-field information at one wavelength to far-field information of a correlated photon with a different wavelength generated through spontaneous photon-pair generation. At the same time, this scheme allows for retrieval of high-contrast images with zero background, making it a highly sensitive scheme for imaging of small objects at challenging spectral ranges with subdiffraction resolutions.

%We use the Green's function quantization method to model the information transfer of an organic molecule with a strong response in the near infrared to its image with the signal photons in the visible range.
\end{abstract}

\maketitle

%\tableofcontents

Quantum imaging with undetected photons (QIUP) works by detecting photons that never interacted with an object \cite{lemos, lahiri,Basset,Kviatkovsky}, using correlation of photon pairs generated through spontaneous parametric down-conversion (SPDC) \cite{Mandel}. Through SPDC in a material with $\chi^{(2)}$-nonlinearity, a pump (P) photon of vacuum wavelength $\lambda_\P$ can split into a pair of signal (S) and idler (I) photons, such that $1/\lambda_\P=1/\lambda_\S+1/\lambda_\I$, where $\lambda_\I$ and $\lambda_\S$ can be far apart.
Conventional QIUP relies on induced coherence (IC) without induced emission, where the object is placed in the path of the idler beam between two identical photon-pair sources \cite{mandel1,mandel2}, typically based on bulk $\chi^{(2)}$-crystals.
Position \cite{vishy,sven} or momentum \cite{lemos,lahiri,lahiri2} correlations between the signal and idler photons can be exploited to construct the object's image by only imaging the signal photons that never interacted with the object.
Hence, object's optical properties at $\lambda_\I$, e.g. its absorption profile, can be inferred in challenging spectral ranges of mid-infrared (MIR) or terahertz (THz), using only detectors in the visible range for $\lambda_\S$ \cite{Kviatkovsky, vanselow, kutas}.

Conventional QIUP relies on far-field interactions, involving only propagating modes, as both object and camera are placed away from the SPDC sources. We use the term ``far-field" in contrast to ``near-field", where near-field interactions can involve evanescent modes and are usually only effective within wavelength-range distances \cite{novotny}.
Hence, resolution of far-field QIUP will be constrained by the diffraction limit \cite{, A&E,Kviatkovsky,fuenza,vishy2}.
For the case where the probing wavelength $\lambda_\I$ is larger than the detection wavelength $\lambda_\S$, object's information is transferred by propagating idler fields and consequently the transverse spatial resolution in this information is constrained by the idler diffraction limit \cite{A&E}.
Hence, conventional QIUP can provide the object's MIR image by only detecting visible photons, avoiding inefficient and expensive MIR detectors, but no resolution advantage is achieved compared to conventional MIR imaging, since the resolution is restricted by the larger MIR wavelength.

To reach subdiffraction resolution, methods like scanning near-field optical microscopy (SNOM) \cite{Chen2012,dai2014,woessner2015,Dai2018, Najmeh}, based on scattering of evanescent fields into propagating fields using a sharp tip in the near field, and nonlinear near-field optical microscopy (NNOM) \cite{frischwasser2021}, based on frequency up-conversion of evanescent fields into higher frequency propagating fields, have been used. However, SNOM techniques require both a source of light and detection at the imaging wavelength, with the additional complication that the tip itself strongly perturbs the field around the object. 
The recently proposed NNOM technique \cite{frischwasser2021} removes the need for detection at the imaging wavelength, yet still requires a source at the wavelength that excites the evanescent fields of the system. This requires elaborate coupling schemes for fields with large in-plane wave-vectors, e.g. plasmonic modes in a metal \cite{frischwasser2021} or a 2D-material layer \cite{dai2014,woessner2015}, where the coupling scheme could itself perturb the system.

\begin{figure}[t]
\includegraphics[scale=1]{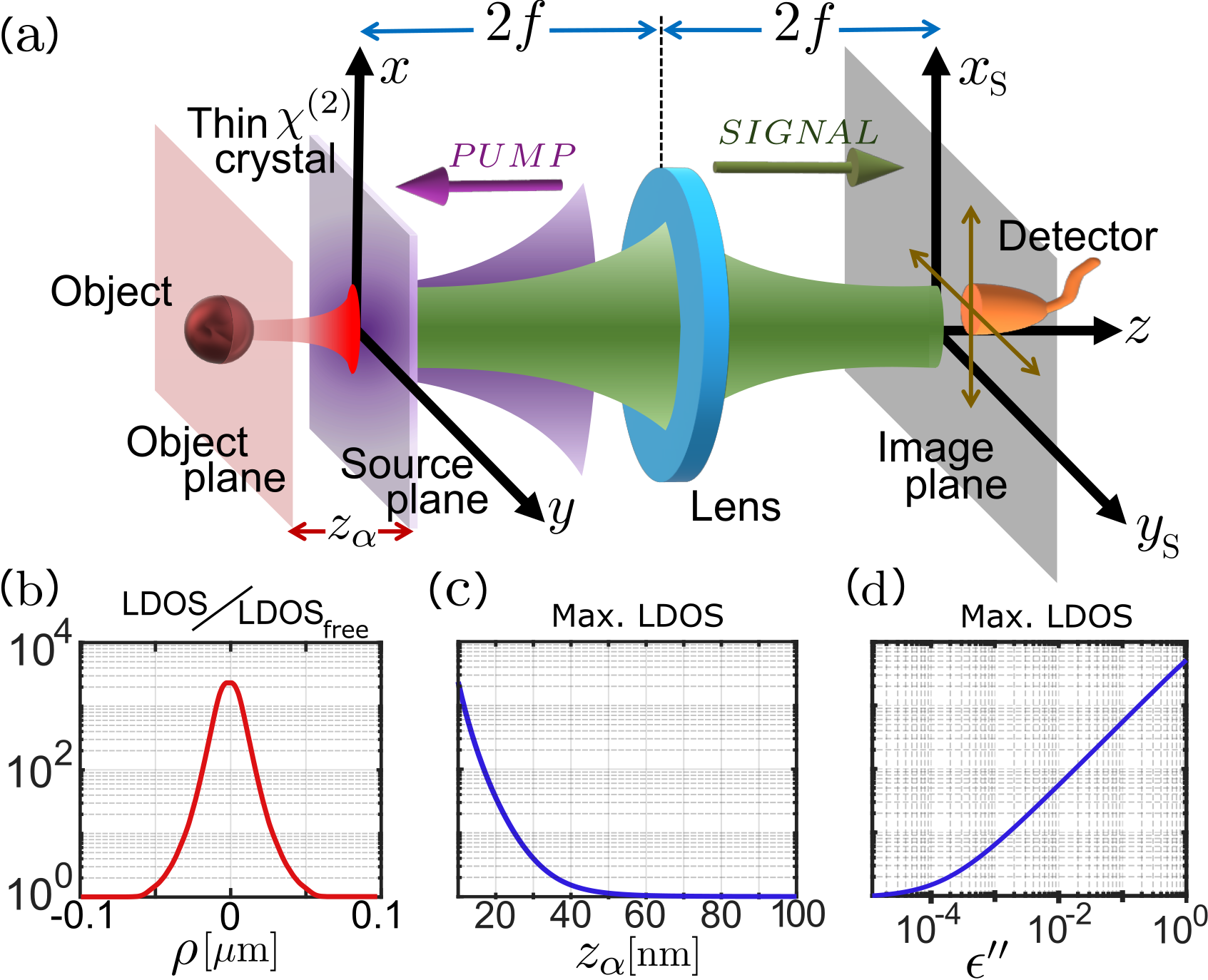}
\caption{\label{fig:sketch} (a) Schematic of the proposed imaging scheme. A thin nonlinear source of photon pairs is illuminated by the pump beam. A nanoparticle interacts with the near field of the source at the idler wavelength, while the signal photons are imaged in the far field. (b) The change in LDOS at $\lambda_\I$ in the source plane, caused by an absorptive nanoparticle at the distance $z_\alpha=10$ nm, where $\rho=\sqrt{x^2+y^2}$ is the source-plane coordinate. The LDOS maximum as a function of nanoparticle's (c) distance to the source $z_\alpha$ and (d) absorption $\epsilon''$. }
\end{figure}

In this work, we propose a subdiffraction imaging scheme based on novel properties of near-field quantum nonlinear interactions, without needing a source or detection at the imaging wavelength, that can access subdiffraction information without needing external near-field probes.
This is done by placing the object in the near field of a planar SPDC source, as shown schematically in Fig.~\ref{fig:sketch}(a).
Whereas conventional QIUP is diffraction-limited by the longer-wavelength idler field that interacts with the object, we numerically verify that our scheme is only diffraction-limited by the shorter-wavelength signal photons that are detected. This makes our scheme subdiffraction with respect to the longer idler wavelength that probes the object.
Moreover, based on the fundamental difference between SPDC with and without near-field interactions, we show how the background photons generated in the absence of the near-field object can be completely removed through angular filtering, allowing for retrieval of high-contrast images with subdiffraction resolution.

Our proposed near-field QIUP is schematically shown in Fig.~\ref{fig:sketch}(a), where the object is placed in the near field of a planar and very thin SPDC source, such that it can directly change the optical properties of the thin source at $\lambda_\I$, while remaining invisible to the signal and pump fields. This could be the case for a small particle that is highly absorptive at $\lambda_\I$ and transparent at $\lambda_\S$ and $\lambda_\P$, e.g. a single molecule with an absorption line at $\lambda_\I$.
The source is then excited by a monochromatic pump beam at $\lambda_\P$.
The signal photons, spectrally filtered to $\lambda_\S=(\lambda_\P^{-1}-\lambda_\I^{-1})^{-1}$, with $\lambda_\S<\lambda_\I$, are collected with an imaging system in the far field.
In this way, the near-field information about the object, which is imprinted onto the source properties at $\lambda_\I$, is transferred to the propagating signal photons at $\lambda_\S$ generated through SPDC.
The use of a thin source is not a prerequisite for observing this effect, however, as it will be explained, has practical importance in making it more pronounced.
In practice, this could be a thin slab of a $\chi^{(2)}$-crystal like lithium niobate \cite{okoth1,santiago1,okoth2}, the surface-nonlinearity of a metal like gold \cite{wang2009}, or a 2D nonlinear material like molybdenum disulfide \cite{saleh}. 
The prospect of such a realization is increased, given the recent observations of pair generation in thin nonlinear crystals and metasurfaces \cite{okoth1,santiago1,okoth2,Anita}.
In this setting, the object can be placed and the signal photons can be detected on either side of the source, because a thin source is not restricted by the phase-matching condition to a forward or backward emission directionality \cite{saleh}. 

For an analytical description, we use the Green's function (GF) description of SPDC, that gives the count rate of signal photons at position $\mathbf{r}_\S$ with frequency $\omega_\S$ and polarization along the unit-vector $\mathbf{d}$  \cite{poddubny,SaraviThesis}:
\begin{align}\label{eq:full_p_S}
    \nonumber & \mathcal{R}_\S  (\mathbf{r}_\S) \propto \sum_{\sigma,\sigma'}d_{\sigma}d^*_{\sigma'}\sum_{\alpha,\beta}\sum_{\alpha',\beta'}\int\dd\mathbf{r}\int\dd\mathbf{r}'\,
    %\\\nonumber&\times
    \Gamma_{\alpha\beta} (\mathbf{r})
    \Gamma_{\alpha'\beta'}^* (\mathbf{r'})
    \\&\times
    \mathrm{Im}[G_{\beta\beta'}(\mathbf{r,r}',\omega_\I)]
    %\\&\times
    G_{\sigma\alpha}(\mathbf{r_\S,r},\omega_\S)G^*_{\sigma'\alpha'}(\mathbf{r_\S,r'},\omega_\S)\, ,
\end{align}
with $\Gamma_{\alpha\beta} (\mathbf{r}) \equiv \sum_\gamma \chi^{(2)}_{\alpha\beta\gamma}(\mathbf{r}) E_{\P,\gamma}(\mathbf{r})$, where $\alpha$, $\beta$, and $\gamma$ indices run over the $x$,$y$, and $z$ directions. Here, $d_{\sigma}$ are the components of $\mathbf{d}$, where $\sigma=x,y,z$.
$E_{\P,\gamma}(\mathbf{r})$ are the vector-components of the complex-valued monochromatic pump with frequency $\omega_\P$, which fixes $\omega_\I=\omega_\P-\omega_\S$. $G_{ij}(\mathbf{r,r'},\omega)$ are the tensor-components of the electric GF. 
Finally, $\chi^{(2)}_{\alpha\beta\gamma}(\mathbf{r})$ are the components of the second-order nonlinear tensor. We take this to be $\chi^{(2)}_{\alpha\beta\gamma}\delta(z)$, describing a 2D planar source at position $z=0$.
In Eq.~(\ref{eq:full_p_S}), system's properties at $\omega_\I$ are captured by the imaginary part of its GF at $\omega_\I$, $\mathrm{Im}[G_{\beta\beta'}(\mathbf{r,r}',\omega_\I)]$. This is the partial cross density of states (CDOS) and describes the spatial coherence between two points at frequency $\omega_\I$ \cite{caze, sauvan, carminati2015}. The local density of states (LDOS) is a special case of CDOS with $\mathbf{r}=\mathbf{r}'$.
IC with two photon-pair sources can more generally be understood as induced changes in CDOS between the nonlinear parts of the system \cite{pawan}.

In our scheme, object's presence in the near field of the source modifies the CDOS within the source, which changes the overall probability and pattern of signal-photon emission.
The idler GF in the object's presence is 
%\begin{align}\label{eq:GF_I}
$\tens{G}(\mathbf{r,r'},\omega_\I) =\tens{G}_0(\mathbf{r,r'},\omega_\I) + \tens{G}_{\rm{sca}} (\mathbf{r,r'},\omega_\I)$,
%\end{align}
where $\tens{G}_0$ is the GF of the free-space background in our case, and describes the system without the object. Here we assume that the 2D nonlinear source has negligible effect on the background GF. $\tens{G}_{\rm{sca}}$ includes the object's effect. Here we take a spherical nanoparticle of radius $a=5$~nm $\ll \lambda_\I$ as the object, which can be approximated as a point scatterer \cite{Castanie, vries,carminati2006,carminati2015}.
For the nanoparticle, we take a material with a strong absorption line in the MIR, namely the compound Diisopropylaminoethanol (DIPA) with relative permittivity $\epsilon=\epsilon'+\ii \epsilon''=1.9763+\ii\,0.39124$ at $\lambda_\I=3.37$ $\mu$m \cite{myers2018accurate}. 
We then analytically find $\tens{G}(\mathbf{r,r'},\omega_\I)$ (see supplemental material \cite{supplement}).
For better visualization, we show the partial LDOS $\mathrm{Im}[G_{xx}(\rho,\rho,\omega_\I)]$ in Fig.~\ref{fig:sketch}(b), normalized to the free-space value, as a function of the radial distance $\rho=\sqrt{x^2+y^2}$ in the source plane, where $\chi^{(2)}\neq 0$. The nanoparticle is centered at $\rho=0$ and $z_\alpha=10$~nm away from the source.
We see a strong and highly spatially localized change of LDOS in the presence of the absorptive nanoparticle, which decays rapidly in the source plane towards the free-space value after about 50~nm. Moreover, the strength of this LDOS change reduces rapidly as the nanoparticle is moved away from the source, as can be seen in Fig.~\ref{fig:sketch}(c).
In Fig.\ref{fig:sketch}(d), we plot the maximum LDOS as a function of
$\epsilon''$ of the nanoparticle's material.
The enhancement of LDOS with loss shows that our proposed effect can in principle be used for sensing the presence of loss at the idler wavelength in any absorptive material system, while at the same time, it shows that a transparent nanoparticle will have negligible contributions to our effect.
Overall, such localized LDOS enhancements are a well-known effect in plasmonics \cite{Castanie,Colasdes,Vandenbem, Vielma}.

Based on Eq.~(\ref{eq:full_p_S}), this change at $\lambda_\I$ manifests itself in the properties of the generated signal photons.
To see this, we consider an imaging system for the signal photons with a magnification of $M=1$ and numerical aperture of $\mathrm{NA}=1$ to evaluate the capability of our scheme at its limit.
This is represented in Fig.~\ref{fig:sketch}(a) with a perfect lens of focal length $f$ placed at a distance $2f$ from the source and the image plane.
To obtain the desired signal GF, we expand the free-space GF at $\omega_\S$ into its spatial-frequency components and keep only propagating waves (see supplemental material \cite{supplement}), which can then be numerically evaluated.
%Hence, signal-photon detection at the image plane will be equivalent to detecting only the propagating components of the generated signal photons directly at the source plane.
%Consequently, we can substitute the signal GF in Eq.~(\ref{eq:full_p_S}) with its propagating part evaluated at the source plane with $z_\S=z=z'=0$.
Without loss of generality, we consider $\chi^{(2)}_{xxx}$ as the dominant component of $\chi^{(2)}_{\alpha\beta\gamma}$. 
We take an $x$-polarized pump beam of width $w=5\;\mu$m at $\lambda_\P=500$~nm, with $E_{\P,x}(x,y)=e^{-(x^2+y^2)/w^2}$, centered on the nanoparticle.
We are interested in imaging the object's properties at $\lambda_\I=3.37$~$\mu$m, which fixes the signal wavelength to $\lambda_\S=587$~nm.
We calculate the unpolarized signal-detection rate by direct summation of Eq.~(\ref{eq:full_p_S}) over three independent polarization directions of the detector, $\mathbf{d}=\hat{x}$, $\hat{y}$, and $\hat{z}$.
Since the idler's GF can be split in two parts, the total detection rate can be split into $\mathcal{R}_\S=\mathcal{R}_0+\mathcal{R}_{\rm{IC}}$:
$\tens{G}_0$ contributes to the background rate, $\mathcal{R}_0$, of detecting signal photons generated by SPDC without the nanoparticle, and $\tens{G}_{\rm{sca}}$ includes the effect of the nanoparticle on signal-photon generation, which we call the induced coherence rate $\mathcal{R}_{\rm{IC}}$.
We first concentrate on $\mathcal{R}_{\rm{IC}}$, and consider the background effect at a later stage.

\begin{figure}[t]
\includegraphics[scale=1]{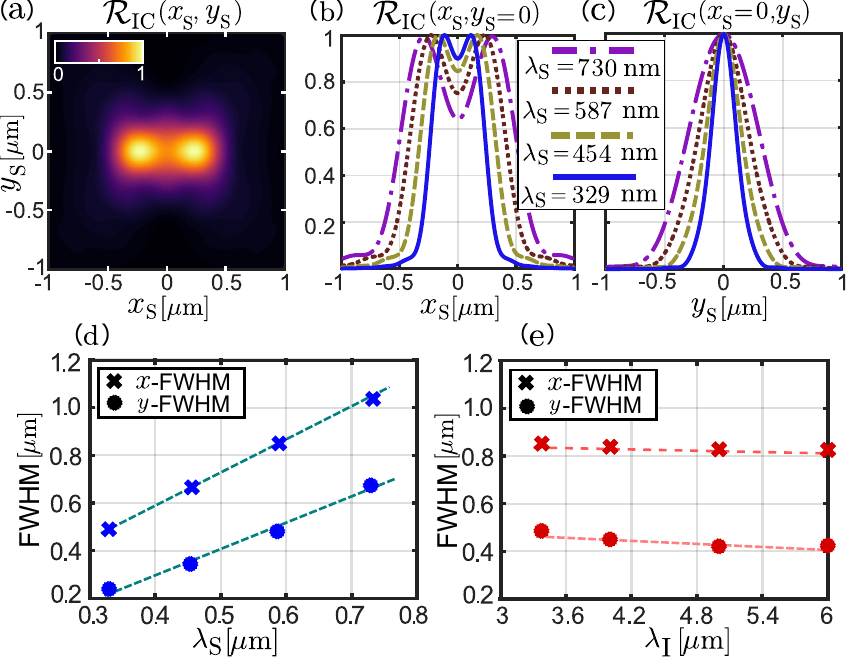}
\caption{\label{fig:P_S} (a) Intensity pattern of the signal photons at $\lambda_\S=587$~nm, caused by the presence of an absorbing nanoparticle at $\lambda_\I=3.37$~$\mu$m, indicating the point-spread function (PSF) of our system. (b) and (c) show the cuts along the $x$ and $y$ axis through the PSF, respectively, for different values of $\lambda_\S$, while keeping $\lambda_\I=3.37$~$\mu$m constant. The corresponding $\lambda_\P$ are $600$~nm (dash-dotted), $500$~nm (dotted), $400$~nm (dashed) and $300$~nm (solid). (d) and (e) show respectively the dependence of the FWHM as a function of $\lambda_S$ with $\lambda_\I=3.37$~$\mu$m constant and as a function of $\lambda_\I$ with $\lambda_\S=587$~nm constant.}
\end{figure}

We show the calculated $\mathcal{R}_{\rm{IC}}(x_\S,y_\S)$ in Fig.~\ref{fig:P_S}(a), with the nanoparticle placed at $z_\alpha=10$~nm. We can think of this pattern as the point-spread function (PSF) of our system, given that we are imaging a point-like object.
Similar to linear imaging, the particular shape of PSF will depend on the detection polarization and the dipolar orientation/shape of the object \cite{Dickson,Lieb,novotny}, where in the supplementary we inspect the role of some of these effects on our PSF shape \cite{supplement}.
More important is the full width at half maximum (FWHM) of the PSF, which indicates our resolution.
Figs.~\ref{fig:P_S}(b) and (c) show cuts of the PSF along the $x$- and $y$-axis, respectively, where the brown dotted lines correspond to the PSF in Fig.~\ref{fig:P_S}(a). 
The FWHM of these curves are $x$-FWHM$\approx 840$~nm and $y$-FWHM$\approx 480$~nm, which are strongly subwavelength values compared to $\lambda_\I=3.37$~$\mu$m.

To understand the wavelength-dependence of resolution, we calculate the PSF cuts for different values of $\lambda_\S$, while keeping $\lambda_\I=3.37$~$\mu$m constant, all shown in Figs.~\ref{fig:P_S}(b,c), and their corresponding FWHMs are plotted in Fig.~\ref{fig:P_S}(d). We do the same calculation by varying $\lambda_\I$ while keeping $\lambda_\S=587$~nm constant, where we only show the resulting FWHMs in Fig.~\ref{fig:P_S}(e). In this case, we assumed the same permittivity at different $\lambda_\I$.
The PSF widths show almost a linear decrease by decreasing $\lambda_\S$, whereas they stay virtually unchanged by varying $\lambda_\I$, showing that indeed the diffraction limit at the shorter $\lambda_\S$ wavelength determines the resolution in our scheme. This behaviour matches our intuitive understanding: the near-field information at $\lambda_\I$ with sub-hundred-nanometer resolution, as seen from the width of LDOS in Fig.~\ref{fig:sketch}(b), is imprinted on the SPDC process with negligible dependence on $\lambda_\I$, and the only limit in transferring this information to the far field is the diffraction limit at $\lambda_\S$.

\begin{figure}[t]
\includegraphics[scale=1]{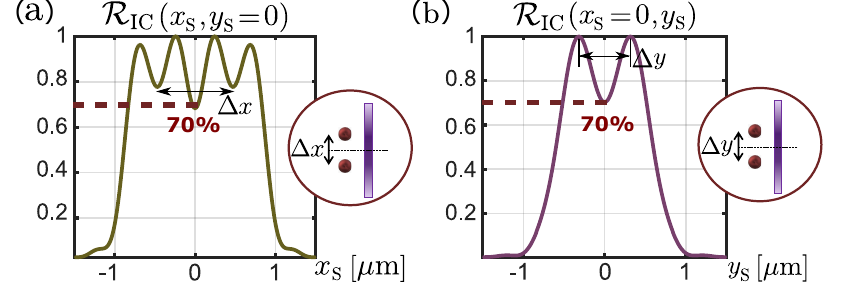}
\caption{\label{fig:two_reso}
The image of two nanoparticles separated by (a) $\Delta x\approx 930$ nm along the $x$-axis and (b) $\Delta y\approx 642$ nm along the $y$-axis, with $\lambda_\S=587$~nm and $\lambda_\I=3.37$~$\mu$m. The distance of the nanoparticles to the source plane is 10~nm and the pump beam is centered at the middle point between them.}
\end{figure}

To verify the resolving power, we consider two nanoparticles at $z_\alpha = 10$~nm and separated from each other along the $x$- or $y$-direction. 
We model this by adding up the $\tens{G}_{\rm{sca}}$ of the two nanoparticles, assuming that they are separated enough that their interaction is negligible (see \cite{supplement} section S3 for a discussion on this approximation).
Since Eq.(\ref{eq:full_p_S}) is linear with respect to idler GF, the PSFs of the two nanoparticles simply add up. This is similar to linear imaging with incoherent illumination \cite{born2013principles}.
We assume that the nanoparticles can be distinguished if the minimum of their PSF overlap is at most $70\%$ of their maximum intensity.
This corresponds to the signal patterns shown in Figs.~\ref{fig:two_reso}(a) and (b), calculated for $\lambda_\S=587$~nm and $\lambda_\I=3.37$~$\mu$m. 
We find the resolution along the $x$- and $y$-directions to be $\Delta x=930$~nm$\approx 1.6\lambda_S$ and $\Delta y=642$~nm$\approx 1.1\lambda_S$, respectively. The linear dependency of resolution on $\lambda_S$ is predicted given the linear dependency of FWHMs on $\lambda_S$, as shown in Fig.~\ref{fig:P_S}(d). In comparison, the diffraction-limited resolution of incoherent linear imaging \cite{born2013principles} is around
$0.6\lambda_\I\approx 2$~$\mu$m. Hence, our achieved resolution in both separation directions is subdiffraction compared to that of linear far-field imaging. Since the resolution in our scheme is mainly restricted by the signal wavelength, the resolution enhancement is even larger for longer idler wavelengths.

\begin{figure}[t]
\includegraphics[scale=1]{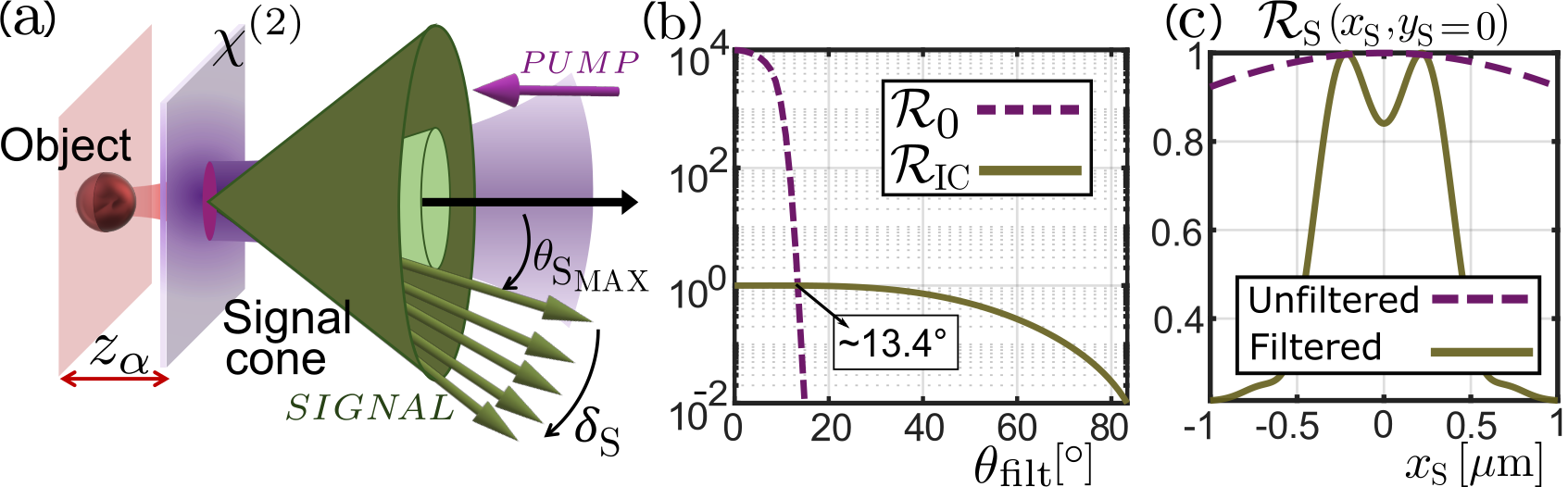}
\caption{\label{fig:filter} (a) Schematic of signal angular-emission range with and without the object in the near-field of the SPDC source. (b) Background and object's contribution to signal intensity, $\mathcal{R}_0$ and $\mathcal{R}_{\rm{IC}}$, at $x_\S=y_\S=0$ as a function of high-pass filtering angle $\theta_\mathrm{filt}$. We normalize the results such that $\mathcal{R}_{\rm{IC}}(\theta_\mathrm{filt}=0,x_\S=y_\S=0)=1$. (c) Total image of the object, unfiltered (dashed line) and after $\theta_{\rm{filt}}\approx 14.3$° filtering (solid line).}
\end{figure}

Up to now, we only showed the object's contribution, $\mathcal{R}_{\rm{IC}}$. We now discuss the background contribution, $\mathcal{R}_0$, and show a method to completely remove it, which also sheds more light on the physics of our scheme.
For a nonlinear crystal much wider than the pump beam, a perfect transverse phase-matching condition is satisfied between the spatial-frequency components of the fields, namely $k_{x\P}=k_{x\S}+k_{x\I}$ and $k_{y\P}=k_{y\S}+k_{y\I}$ \cite{howell}. Since there is no longitudinal phase-matching restriction in a very thin source, propagating signal and idler photons can be generated in any pairs of angular directions, as long as they satisfy the transverse phase-matching condition \cite{saleh,okoth1,santiago1,okoth2}.
For a normally incident plane-wave pump, with $k_{x\P}=k_{y\P}=0$, the emitted pair satisfy $k_{x\I}=-k_{x\S}$ and $k_{y\I}=-k_{y\S}$. Considering $\theta$ as the emission angle with respect to the forward $z$-direction we get $k_{x}^2+k_{y}^2=k^2\sin^2\theta$.
Defining $r\equiv\lambda_\S/\lambda_\I$ as the degeneracy factor, we get $|\sin\theta_\S|=r|\sin\theta_\I|$ for the emission angles.
%, where all possible combinations of forward and backward emission are included.
For $r<1$, the maximum signal emission angle is $|\theta_{\S_\mathrm{MAX}}|=| \sin^{-1} (r) |$.
Hence, when $\lambda_\I>\lambda_\S$, idler photons are emitted in a full angular range, while signal photons have a reduced degeneracy-dependent range.
With $\lambda_\I=3.37$~$\mu$m and $\lambda_\S=587$~nm, we have $r=0.174$, which corresponds to $\theta_{\S_\mathrm{MAX}}\approx \pm 10$°.
%, whereas $\theta_{\I_\mathrm{MAX}}= \pm 90$°.
%while the idler photons can be generated in the full range of $\pm 90$°. 
The angular emission cone is shown schematically in Fig.~\ref{fig:filter}(a) for signal photons propagating in the forward $z$-direction.

Notably, evanescent idler modes with imaginary-valued $k_z$ do not participate in background's contribution. This was shown in periodic structures, where bandgap modes with zero DOS do not contribute to pair generation, and the presence of a 2-level system was used to add DOS in the bandgap and mediate SPDC \cite{sina}.
The nanoparticle's presence in the near field of the source in our scheme has a similar effect, as it involves the evanescent idler modes with $k_x^2+k_y^2>k_\I^2$ in SPDC and results in signal generation in a larger angular range, marked schematically by $\delta_\S$ in Fig.~\ref{fig:filter}(a).
The different angular ranges for the background and object's contributions allows for filtering out the background completely, while still retaining the object's contribution, enhancing the image contrast.

We verify this by calculating $\mathcal{R}_0$ and $\mathcal{R}_{\rm{IC}}$, when signal photons are collected after a filter that blocks angular components with $\theta<\theta_\mathrm{filt}$. In practice, this could be done using a 4-f setup \cite{saleh2019fundamentals}.
The filtered $\mathcal{R}_0$ and $\mathcal{R}_{\rm{IC}}$ at the detection point $x_\S=y_\S=0$ are shown in Fig.~\ref{fig:filter}(b) as a function of filtering angle $\theta_\mathrm{filt}$. The background contribution decreases strongly around $\theta_\mathrm{filt}=10$° and after $\theta_{\rm{filt}}\approx 13.4$° becomes smaller than the object's contribution. In Fig.~\ref{fig:filter}(c), we show the full $\mathcal{R}_\S=\mathcal{R}_0+\mathcal{R}_{\rm{IC}}$ for the unfiltered and filtered at $\theta_{\rm{filt}}\approx 14.3$° case,
showing that we can retrieve the PSF shape after filtering.
The fact that $\mathcal{R}_0$ does not fully vanish at $\theta_{\rm{filt}}\approx 10$°, is because of the finite-sized pump beam, which contains transverse spatial-frequency components that broaden the phase-matching condition for the background SPDC.
Notably, $\mathcal{R}_{\rm{IC}}$ barely changes when filtering out $\mathcal{R}_0$, showing that a large part of object's information is embedded in the large-angle signal photons. This verifies that evanescent idler modes with transverse wave-vectors larger than $k_\I$ and up to $k_\S$ are involved in object's contribution to SPDC.
We note that a thin source serves to minimize $\mathcal{R}_0$, as the object's DOS enhancement is only effective within tens of nanometers away from it and a thicker source potentially only adds to the background.

Finally, we calculate the total signal photon generation rate associated to the presence of the object, based on the available GF-based formulations \cite{poddubny,SaraviThesis}. To do this, we consider a 10~nm thick slab of gallium phosphide (GaP) \cite{shoji}, as an approximation for a 2D nonlinear source. We consider a $x$-polarized pump, which results in cross-polarized signal photons with $y$ and $z$ polarizations. We consider detection over the range of $\lambda_\S$ from $577$ to $591$ nm, which corresponds to probing the $\lambda_\I$ range from $3.24$ to $3.75$ $\mu$m that encompasses an absorption line of DIPA. We consider three different DIPA nanoparticles with radii of 5, 10, and 50~nm, all placed 5 nm away from the surface of the nonlinear material to the bottom of the spherical nanoparticle. With a pump power of 100~mW, spread over a Gaussian beam of $w=5\;\mu$m, we obtain a total generation rate of $0.06,\; 0.43,\; 25.66$ signal photons per second for the three nanoparticle sizes, respectively. The details of our calculation are included in the supplementary \cite{supplement}. Such signals can in principle be measured using state-of-the-art cooled CCD cameras.

Overall, our studies show, that the generated signal photons have signature properties in the spectral, spatial, and polarization degrees of freedom, which strongly depend on the properties of the system and the nanoparticle itself. This can be used for both filtering out unwanted contributions, but also to detect the properties of the nano-object, e.g. its absorption spectrum and shape/orientation.

Importantly, our scheme can also be used in a "classical" regime, by seeding at $\lambda_\S$ and looking at the seed's intensity change caused by difference-frequency generation with the pump beam, which probes the DOS at $\lambda_\I$. This however loses the zero-background advantage, as we have to seed the signal in the same angular range where we detect, creating an unavoidable background.

In conclusion, we proposed and numerically investigated a near-field QIUP scheme with subdiffraction resolution and zero background. Potential applications are in probing the structure of highly confined polariton modes in structures made of 2D and van der Waals materials \cite{Chen2012,dai2014,woessner2015,Dai2018}, infrared nanoscopy of strained semiconductors \cite{huber2009infrared}, absorption-based detection of single non-fluorescent molecules \cite{Celebrano2011}, and in general optical detection of weakly scattering systems, like nanoparticles, molecules, or biological samples that have their signature absorption lines at spectral ranges like MIR \cite{knoll1999near,mayet2008sub,ballout2011scanning,huth2012nano}. This is all without needing SNOM and without a source or a detector at the mid- and long-IR wavelength ranges, while simultaneously benefiting from the spatial resolution of visible wavelengths.
More generally, our work sheds new light onto the physics of near-field quantum nonlinear interactions in the presence of absorptive systems. This is especially important for creation of hybrid integrated hardware for quantum technologies \cite{Elshaari2020}, in which a nonlinear system with a bosonic field is directly interfaced with a coherently absorbing quantum system, e.g. a single-photon emitter \cite{sina}.

\begin{acknowledgments}This research is supported by the by German Ministry of Education and Research (13N14877) and Free State of Turingia (2017 FRG 0067). We also thank Andres Vega for useful discussions.\end{acknowledgments}

%\nocite{*}


\begin{thebibliography}{56}%
\makeatletter
\providecommand \@ifxundefined [1]{%
 \@ifx{#1\undefined}
}%
\providecommand \@ifnum [1]{%
 \ifnum #1\expandafter \@firstoftwo
 \else \expandafter \@secondoftwo
 \fi
}%
\providecommand \@ifx [1]{%
 \ifx #1\expandafter \@firstoftwo
 \else \expandafter \@secondoftwo
 \fi
}%
\providecommand \natexlab [1]{#1}%
\providecommand \enquote  [1]{``#1''}%
\providecommand \bibnamefont  [1]{#1}%
\providecommand \bibfnamefont [1]{#1}%
\providecommand \citenamefont [1]{#1}%
\providecommand \href@noop [0]{\@secondoftwo}%
\providecommand \href [0]{\begingroup \@sanitize@url \@href}%
\providecommand \@href[1]{\@@startlink{#1}\@@href}%
\providecommand \@@href[1]{\endgroup#1\@@endlink}%
\providecommand \@sanitize@url [0]{\catcode `\\12\catcode `\$12\catcode
  `\&12\catcode `\#12\catcode `\^12\catcode `\_12\catcode `\%12\relax}%
\providecommand \@@startlink[1]{}%
\providecommand \@@endlink[0]{}%
\providecommand \url  [0]{\begingroup\@sanitize@url \@url }%
\providecommand \@url [1]{\endgroup\@href {#1}{\urlprefix }}%
\providecommand \urlprefix  [0]{URL }%
\providecommand \Eprint [0]{\href }%
\providecommand \doibase [0]{http://dx.doi.org/}%
\providecommand \selectlanguage [0]{\@gobble}%
\providecommand \bibinfo  [0]{\@secondoftwo}%
\providecommand \bibfield  [0]{\@secondoftwo}%
\providecommand \translation [1]{[#1]}%
\providecommand \BibitemOpen [0]{}%
\providecommand \bibitemStop [0]{}%
\providecommand \bibitemNoStop [0]{.\EOS\space}%
\providecommand \EOS [0]{\spacefactor3000\relax}%
\providecommand \BibitemShut  [1]{\csname bibitem#1\endcsname}%
\let\auto@bib@innerbib\@empty
%</preamble>
\bibitem [{\citenamefont {Lemos}\ \emph {et~al.}(2014)\citenamefont {Lemos},
  \citenamefont {Borish}, \citenamefont {Cole}, \citenamefont {Ramelow},
  \citenamefont {Lapkiewicz},\ and\ \citenamefont {Zeilinger}}]{lemos}%
  \BibitemOpen
  \bibfield  {author} {\bibinfo {author} {\bibfnamefont {G.~B.}\ \bibnamefont
  {Lemos}}, \bibinfo {author} {\bibfnamefont {V.}~\bibnamefont {Borish}},
  \bibinfo {author} {\bibfnamefont {G.~D.}\ \bibnamefont {Cole}}, \bibinfo
  {author} {\bibfnamefont {S.}~\bibnamefont {Ramelow}}, \bibinfo {author}
  {\bibfnamefont {R.}~\bibnamefont {Lapkiewicz}}, \ and\ \bibinfo {author}
  {\bibfnamefont {A.}~\bibnamefont {Zeilinger}},\ }\href
  {https://www.nature.com/articles/nature13586} {\bibfield  {journal} {\bibinfo
   {journal} {Nature}\ }\textbf {\bibinfo {volume} {512}},\ \bibinfo {pages}
  {409} (\bibinfo {year} {2014})}\BibitemShut {NoStop}%
\bibitem [{\citenamefont {Lahiri}\ \emph {et~al.}(2015)\citenamefont {Lahiri},
  \citenamefont {Lapkiewicz}, \citenamefont {Lemos},\ and\ \citenamefont
  {Zeilinger}}]{lahiri}%
  \BibitemOpen
  \bibfield  {author} {\bibinfo {author} {\bibfnamefont {M.}~\bibnamefont
  {Lahiri}}, \bibinfo {author} {\bibfnamefont {R.}~\bibnamefont {Lapkiewicz}},
  \bibinfo {author} {\bibfnamefont {G.~B.}\ \bibnamefont {Lemos}}, \ and\
  \bibinfo {author} {\bibfnamefont {A.}~\bibnamefont {Zeilinger}},\ }\href
  {\doibase 10.1103/PhysRevA.92.013832} {\bibfield  {journal} {\bibinfo
  {journal} {Physical Review A}\ }\textbf {\bibinfo {volume} {92}},\ \bibinfo
  {pages} {013832} (\bibinfo {year} {2015})}\BibitemShut {NoStop}%
\bibitem [{\citenamefont {Gilaberte~Basset}\ \emph {et~al.}(2021)\citenamefont
  {Gilaberte~Basset}, \citenamefont {Hochrainer}, \citenamefont {Töpfer},
  \citenamefont {Riexinger}, \citenamefont {Bickert}, \citenamefont
  {León-Torres}, \citenamefont {Steinlechner},\ and\ \citenamefont
  {Gräfe}}]{Basset}%
  \BibitemOpen
  \bibfield  {author} {\bibinfo {author} {\bibfnamefont {M.}~\bibnamefont
  {Gilaberte~Basset}}, \bibinfo {author} {\bibfnamefont {A.}~\bibnamefont
  {Hochrainer}}, \bibinfo {author} {\bibfnamefont {S.}~\bibnamefont {Töpfer}},
  \bibinfo {author} {\bibfnamefont {F.}~\bibnamefont {Riexinger}}, \bibinfo
  {author} {\bibfnamefont {P.}~\bibnamefont {Bickert}}, \bibinfo {author}
  {\bibfnamefont {J.~R.}\ \bibnamefont {León-Torres}}, \bibinfo {author}
  {\bibfnamefont {F.}~\bibnamefont {Steinlechner}}, \ and\ \bibinfo {author}
  {\bibfnamefont {M.}~\bibnamefont {Gräfe}},\ }\href {\doibase
  https://doi.org/10.1002/lpor.202000327} {\bibfield  {journal} {\bibinfo
  {journal} {Laser \& Photonics Reviews}\ }\textbf {\bibinfo {volume} {15}},\
  \bibinfo {pages} {2000327} (\bibinfo {year} {2021})}\BibitemShut {NoStop}%
\bibitem [{\citenamefont {{Kviatkovsky}}\ \emph {et~al.}(2020)\citenamefont
  {{Kviatkovsky}}, \citenamefont {{Chrzanowski}}, \citenamefont {{Avery}},
  \citenamefont {{Bartolomaeus}},\ and\ \citenamefont
  {{Ramelow}}}]{Kviatkovsky}%
  \BibitemOpen
  \bibfield  {author} {\bibinfo {author} {\bibfnamefont {I.}~\bibnamefont
  {{Kviatkovsky}}}, \bibinfo {author} {\bibfnamefont {H.~M.}\ \bibnamefont
  {{Chrzanowski}}}, \bibinfo {author} {\bibfnamefont {E.~G.}\ \bibnamefont
  {{Avery}}}, \bibinfo {author} {\bibfnamefont {H.}~\bibnamefont
  {{Bartolomaeus}}}, \ and\ \bibinfo {author} {\bibfnamefont {S.}~\bibnamefont
  {{Ramelow}}},\ }\href {\doibase 10.1126/sciadv.abd0264} {\bibfield  {journal}
  {\bibinfo  {journal} {Science Advances}\ }\textbf {\bibinfo {volume} {6}},\
  \bibinfo {eid} {eabd0264} (\bibinfo {year} {2020})}\BibitemShut {NoStop}%
\bibitem [{\citenamefont {Hong}\ and\ \citenamefont {Mandel}(1985)}]{Mandel}%
  \BibitemOpen
  \bibfield  {author} {\bibinfo {author} {\bibfnamefont {C.~K.}\ \bibnamefont
  {Hong}}\ and\ \bibinfo {author} {\bibfnamefont {L.}~\bibnamefont {Mandel}},\
  }\href {\doibase 10.1103/PhysRevA.31.2409} {\bibfield  {journal} {\bibinfo
  {journal} {Physical Review A}\ }\textbf {\bibinfo {volume} {31}},\ \bibinfo
  {pages} {2409} (\bibinfo {year} {1985})}\BibitemShut {NoStop}%
\bibitem [{\citenamefont {Zou}\ \emph {et~al.}(1991)\citenamefont {Zou},
  \citenamefont {Wang},\ and\ \citenamefont {Mandel}}]{mandel1}%
  \BibitemOpen
  \bibfield  {author} {\bibinfo {author} {\bibfnamefont {X.~Y.}\ \bibnamefont
  {Zou}}, \bibinfo {author} {\bibfnamefont {L.~J.}\ \bibnamefont {Wang}}, \
  and\ \bibinfo {author} {\bibfnamefont {L.}~\bibnamefont {Mandel}},\ }\href
  {\doibase 10.1103/PhysRevLett.67.318} {\bibfield  {journal} {\bibinfo
  {journal} {Physical Review Letters}\ }\textbf {\bibinfo {volume} {67}},\
  \bibinfo {pages} {318} (\bibinfo {year} {1991})}\BibitemShut {NoStop}%
\bibitem [{\citenamefont {Wang}\ \emph {et~al.}(1991)\citenamefont {Wang},
  \citenamefont {Zou},\ and\ \citenamefont {Mandel}}]{mandel2}%
  \BibitemOpen
  \bibfield  {author} {\bibinfo {author} {\bibfnamefont {L.~J.}\ \bibnamefont
  {Wang}}, \bibinfo {author} {\bibfnamefont {X.~Y.}\ \bibnamefont {Zou}}, \
  and\ \bibinfo {author} {\bibfnamefont {L.}~\bibnamefont {Mandel}},\ }\href
  {\doibase 10.1103/PhysRevA.44.4614} {\bibfield  {journal} {\bibinfo
  {journal} {Physical Review A}\ }\textbf {\bibinfo {volume} {44}},\ \bibinfo
  {pages} {4614} (\bibinfo {year} {1991})}\BibitemShut {NoStop}%
\bibitem [{\citenamefont {Viswanathan}\ \emph
  {et~al.}(2021{\natexlab{a}})\citenamefont {Viswanathan}, \citenamefont
  {Barreto~Lemos},\ and\ \citenamefont {Lahiri}}]{vishy}%
  \BibitemOpen
  \bibfield  {author} {\bibinfo {author} {\bibfnamefont {B.}~\bibnamefont
  {Viswanathan}}, \bibinfo {author} {\bibfnamefont {G.}~\bibnamefont
  {Barreto~Lemos}}, \ and\ \bibinfo {author} {\bibfnamefont {M.}~\bibnamefont
  {Lahiri}},\ }\href {\doibase 10.1364/ol.419502} {\bibfield  {journal}
  {\bibinfo  {journal} {Optics Letters}\ }\textbf {\bibinfo {volume} {46}},\
  \bibinfo {pages} {3496} (\bibinfo {year} {2021}{\natexlab{a}})}\BibitemShut
  {NoStop}%
\bibitem [{\citenamefont {Kviatkovsky}\ \emph {et~al.}(2022)\citenamefont
  {Kviatkovsky}, \citenamefont {Chrzanowski},\ and\ \citenamefont
  {Ramelow}}]{sven}%
  \BibitemOpen
  \bibfield  {author} {\bibinfo {author} {\bibfnamefont {I.}~\bibnamefont
  {Kviatkovsky}}, \bibinfo {author} {\bibfnamefont {H.~M.}\ \bibnamefont
  {Chrzanowski}}, \ and\ \bibinfo {author} {\bibfnamefont {S.}~\bibnamefont
  {Ramelow}},\ }\href {\doibase 10.1364/OE.440534} {\bibfield  {journal}
  {\bibinfo  {journal} {Optics Express}\ }\textbf {\bibinfo {volume} {30}},\
  \bibinfo {pages} {5916} (\bibinfo {year} {2022})}\BibitemShut {NoStop}%
\bibitem [{\citenamefont {Lahiri}\ \emph {et~al.}(2017)\citenamefont {Lahiri},
  \citenamefont {Hochrainer}, \citenamefont {Lapkiewicz}, \citenamefont
  {Lemos},\ and\ \citenamefont {Zeilinger}}]{lahiri2}%
  \BibitemOpen
  \bibfield  {author} {\bibinfo {author} {\bibfnamefont {M.}~\bibnamefont
  {Lahiri}}, \bibinfo {author} {\bibfnamefont {A.}~\bibnamefont {Hochrainer}},
  \bibinfo {author} {\bibfnamefont {R.}~\bibnamefont {Lapkiewicz}}, \bibinfo
  {author} {\bibfnamefont {G.~B.}\ \bibnamefont {Lemos}}, \ and\ \bibinfo
  {author} {\bibfnamefont {A.}~\bibnamefont {Zeilinger}},\ }\href {\doibase
  10.1103/PhysRevA.96.013822} {\bibfield  {journal} {\bibinfo  {journal}
  {Physical Review A}\ }\textbf {\bibinfo {volume} {96}},\ \bibinfo {pages}
  {013822} (\bibinfo {year} {2017})}\BibitemShut {NoStop}%
\bibitem [{\citenamefont {Vanselow}\ \emph {et~al.}(2020)\citenamefont
  {Vanselow}, \citenamefont {Kaufmann}, \citenamefont {Zorin}, \citenamefont
  {Heise}, \citenamefont {Chrzanowski},\ and\ \citenamefont
  {Ramelow}}]{vanselow}%
  \BibitemOpen
  \bibfield  {author} {\bibinfo {author} {\bibfnamefont {A.}~\bibnamefont
  {Vanselow}}, \bibinfo {author} {\bibfnamefont {P.}~\bibnamefont {Kaufmann}},
  \bibinfo {author} {\bibfnamefont {I.}~\bibnamefont {Zorin}}, \bibinfo
  {author} {\bibfnamefont {B.}~\bibnamefont {Heise}}, \bibinfo {author}
  {\bibfnamefont {H.~M.}\ \bibnamefont {Chrzanowski}}, \ and\ \bibinfo {author}
  {\bibfnamefont {S.}~\bibnamefont {Ramelow}},\ }\href
  {http://www.osapublishing.org/optica/abstract.cfm?URI=optica-7-12-1729}
  {\bibfield  {journal} {\bibinfo  {journal} {Optica}\ }\textbf {\bibinfo
  {volume} {7}},\ \bibinfo {pages} {1729} (\bibinfo {year} {2020})}\BibitemShut
  {NoStop}%
\bibitem [{\citenamefont {Kutas}\ \emph {et~al.}(2020)\citenamefont {Kutas},
  \citenamefont {Haase}, \citenamefont {Bickert}, \citenamefont {Riexinger},
  \citenamefont {Molter},\ and\ \citenamefont {von Freymann}}]{kutas}%
  \BibitemOpen
  \bibfield  {author} {\bibinfo {author} {\bibfnamefont {M.}~\bibnamefont
  {Kutas}}, \bibinfo {author} {\bibfnamefont {B.}~\bibnamefont {Haase}},
  \bibinfo {author} {\bibfnamefont {P.}~\bibnamefont {Bickert}}, \bibinfo
  {author} {\bibfnamefont {F.}~\bibnamefont {Riexinger}}, \bibinfo {author}
  {\bibfnamefont {D.}~\bibnamefont {Molter}}, \ and\ \bibinfo {author}
  {\bibfnamefont {G.}~\bibnamefont {von Freymann}},\ }\href
  {https://advances.sciencemag.org/content/6/11/eaaz8065} {\bibfield  {journal}
  {\bibinfo  {journal} {Science Advances}\ }\textbf {\bibinfo {volume} {6}},\
  \bibinfo {pages} {eaaz8065} (\bibinfo {year} {2020})}\BibitemShut {NoStop}%
\bibitem [{\citenamefont {Novotny}\ and\ \citenamefont
  {Hecht}(2006)}]{novotny}%
  \BibitemOpen
  \bibfield  {author} {\bibinfo {author} {\bibfnamefont {L.}~\bibnamefont
  {Novotny}}\ and\ \bibinfo {author} {\bibfnamefont {B.}~\bibnamefont
  {Hecht}},\ }\href {\doibase 10.1017/CBO9780511813535} {\emph {\bibinfo
  {title} {Principles of Nano-Optics}}}\ (\bibinfo  {publisher} {Cambridge
  University Press},\ \bibinfo {year} {2006})\BibitemShut {NoStop}%
\bibitem [{\citenamefont {Vega}\ \emph {et~al.}(2022)\citenamefont {Vega},
  \citenamefont {Santos}, \citenamefont {Fuenzalida}, \citenamefont {Basset},
  \citenamefont {Pertsch}, \citenamefont {Grafe}, \citenamefont {Saravi},\ and\
  \citenamefont {Setzpfandt}}]{A&E}%
  \BibitemOpen
  \bibfield  {author} {\bibinfo {author} {\bibfnamefont {A.}~\bibnamefont
  {Vega}}, \bibinfo {author} {\bibfnamefont {E.~A.}\ \bibnamefont {Santos}},
  \bibinfo {author} {\bibfnamefont {J.}~\bibnamefont {Fuenzalida}}, \bibinfo
  {author} {\bibfnamefont {M.~G.}\ \bibnamefont {Basset}}, \bibinfo {author}
  {\bibfnamefont {T.}~\bibnamefont {Pertsch}}, \bibinfo {author} {\bibfnamefont
  {M.}~\bibnamefont {Grafe}}, \bibinfo {author} {\bibfnamefont
  {S.}~\bibnamefont {Saravi}}, \ and\ \bibinfo {author} {\bibfnamefont
  {F.}~\bibnamefont {Setzpfandt}},\ }\href@noop {} {\enquote {\bibinfo {title}
  {Fundamental resolution limit of quantum imaging with undetected photons},}\
  } (\bibinfo {year} {2022}),\ \Eprint {http://arxiv.org/abs/arXiv:2203.06106}
  {arXiv:2203.06106} \BibitemShut {NoStop}%
\bibitem [{\citenamefont {Fuenzalida}\ \emph {et~al.}(2022)\citenamefont
  {Fuenzalida}, \citenamefont {Hochrainer}, \citenamefont {Lemos},
  \citenamefont {Ortega}, \citenamefont {Lapkiewicz}, \citenamefont {Lahiri},\
  and\ \citenamefont {Zeilinger}}]{fuenza}%
  \BibitemOpen
  \bibfield  {author} {\bibinfo {author} {\bibfnamefont {J.}~\bibnamefont
  {Fuenzalida}}, \bibinfo {author} {\bibfnamefont {A.}~\bibnamefont
  {Hochrainer}}, \bibinfo {author} {\bibfnamefont {G.~B.}\ \bibnamefont
  {Lemos}}, \bibinfo {author} {\bibfnamefont {E.~A.}\ \bibnamefont {Ortega}},
  \bibinfo {author} {\bibfnamefont {R.}~\bibnamefont {Lapkiewicz}}, \bibinfo
  {author} {\bibfnamefont {M.}~\bibnamefont {Lahiri}}, \ and\ \bibinfo {author}
  {\bibfnamefont {A.}~\bibnamefont {Zeilinger}},\ }\href {\doibase
  10.22331/q-2022-02-09-646} {\bibfield  {journal} {\bibinfo  {journal}
  {{Quantum}}\ }\textbf {\bibinfo {volume} {6}},\ \bibinfo {pages} {646}
  (\bibinfo {year} {2022})}\BibitemShut {NoStop}%
\bibitem [{\citenamefont {Viswanathan}\ \emph
  {et~al.}(2021{\natexlab{b}})\citenamefont {Viswanathan}, \citenamefont
  {Lemos},\ and\ \citenamefont {Lahiri}}]{vishy2}%
  \BibitemOpen
  \bibfield  {author} {\bibinfo {author} {\bibfnamefont {B.}~\bibnamefont
  {Viswanathan}}, \bibinfo {author} {\bibfnamefont {G.~B.}\ \bibnamefont
  {Lemos}}, \ and\ \bibinfo {author} {\bibfnamefont {M.}~\bibnamefont
  {Lahiri}},\ }\href {\doibase 10.1364/OE.434085} {\bibfield  {journal}
  {\bibinfo  {journal} {Optics Express}\ }\textbf {\bibinfo {volume} {29}},\
  \bibinfo {pages} {38185} (\bibinfo {year} {2021}{\natexlab{b}})}\BibitemShut
  {NoStop}%
\bibitem [{\citenamefont {Chen}\ \emph {et~al.}(2012)\citenamefont {Chen},
  \citenamefont {Badioli}, \citenamefont {Alonso-Gonz{\'a}lez}, \citenamefont
  {Thongrattanasiri}, \citenamefont {Huth}, \citenamefont {Osmond},
  \citenamefont {Spasenovi{\'c}}, \citenamefont {Centeno}, \citenamefont
  {Pesquera}, \citenamefont {Godignon} \emph {et~al.}}]{Chen2012}%
  \BibitemOpen
  \bibfield  {author} {\bibinfo {author} {\bibfnamefont {J.}~\bibnamefont
  {Chen}}, \bibinfo {author} {\bibfnamefont {M.}~\bibnamefont {Badioli}},
  \bibinfo {author} {\bibfnamefont {P.}~\bibnamefont {Alonso-Gonz{\'a}lez}},
  \bibinfo {author} {\bibfnamefont {S.}~\bibnamefont {Thongrattanasiri}},
  \bibinfo {author} {\bibfnamefont {F.}~\bibnamefont {Huth}}, \bibinfo {author}
  {\bibfnamefont {J.}~\bibnamefont {Osmond}}, \bibinfo {author} {\bibfnamefont
  {M.}~\bibnamefont {Spasenovi{\'c}}}, \bibinfo {author} {\bibfnamefont
  {A.}~\bibnamefont {Centeno}}, \bibinfo {author} {\bibfnamefont
  {A.}~\bibnamefont {Pesquera}}, \bibinfo {author} {\bibfnamefont
  {P.}~\bibnamefont {Godignon}},  \emph {et~al.},\ }\href
  {https://doi.org/10.1038/nature11254} {\bibfield  {journal} {\bibinfo
  {journal} {Nature}\ }\textbf {\bibinfo {volume} {487}},\ \bibinfo {pages}
  {77} (\bibinfo {year} {2012})}\BibitemShut {NoStop}%
\bibitem [{\citenamefont {{Dai}}\ \emph {et~al.}(2014)\citenamefont {{Dai}},
  \citenamefont {{Fei}}, \citenamefont {{Ma}}, \citenamefont {{Rodin}},
  \citenamefont {{Wagner}}, \citenamefont {{McLeod}}, \citenamefont {{Liu}},
  \citenamefont {{Gannett}}, \citenamefont {{Regan}}, \citenamefont
  {{Watanabe}}, \citenamefont {{Taniguchi}}, \citenamefont {{Thiemens}},
  \citenamefont {{Dominguez}}, \citenamefont {{Neto}}, \citenamefont {{Zettl}},
  \citenamefont {{Keilmann}}, \citenamefont {{Jarillo-Herrero}}, \citenamefont
  {{Fogler}},\ and\ \citenamefont {{Basov}}}]{dai2014}%
  \BibitemOpen
  \bibfield  {author} {\bibinfo {author} {\bibfnamefont {S.}~\bibnamefont
  {{Dai}}}, \bibinfo {author} {\bibfnamefont {Z.}~\bibnamefont {{Fei}}},
  \bibinfo {author} {\bibfnamefont {Q.}~\bibnamefont {{Ma}}}, \bibinfo {author}
  {\bibfnamefont {A.~S.}\ \bibnamefont {{Rodin}}}, \bibinfo {author}
  {\bibfnamefont {M.}~\bibnamefont {{Wagner}}}, \bibinfo {author}
  {\bibfnamefont {A.~S.}\ \bibnamefont {{McLeod}}}, \bibinfo {author}
  {\bibfnamefont {M.~K.}\ \bibnamefont {{Liu}}}, \bibinfo {author}
  {\bibfnamefont {W.}~\bibnamefont {{Gannett}}}, \bibinfo {author}
  {\bibfnamefont {W.}~\bibnamefont {{Regan}}}, \bibinfo {author} {\bibfnamefont
  {K.}~\bibnamefont {{Watanabe}}}, \bibinfo {author} {\bibfnamefont
  {T.}~\bibnamefont {{Taniguchi}}}, \bibinfo {author} {\bibfnamefont
  {M.}~\bibnamefont {{Thiemens}}}, \bibinfo {author} {\bibfnamefont
  {G.}~\bibnamefont {{Dominguez}}}, \bibinfo {author} {\bibfnamefont
  {A.~H.~C.}\ \bibnamefont {{Neto}}}, \bibinfo {author} {\bibfnamefont
  {A.}~\bibnamefont {{Zettl}}}, \bibinfo {author} {\bibfnamefont
  {F.}~\bibnamefont {{Keilmann}}}, \bibinfo {author} {\bibfnamefont
  {P.}~\bibnamefont {{Jarillo-Herrero}}}, \bibinfo {author} {\bibfnamefont
  {M.~M.}\ \bibnamefont {{Fogler}}}, \ and\ \bibinfo {author} {\bibfnamefont
  {D.~N.}\ \bibnamefont {{Basov}}},\ }\href {\doibase 10.1126/science.1246833}
  {\bibfield  {journal} {\bibinfo  {journal} {Science}\ }\textbf {\bibinfo
  {volume} {343}},\ \bibinfo {pages} {1125} (\bibinfo {year}
  {2014})}\BibitemShut {NoStop}%
\bibitem [{\citenamefont {{Woessner}}\ \emph {et~al.}(2015)\citenamefont
  {{Woessner}}, \citenamefont {{Lundeberg}}, \citenamefont {{Gao}},
  \citenamefont {{Principi}}, \citenamefont {{Alonso-Gonz{\'a}lez}},
  \citenamefont {{Carrega}}, \citenamefont {{Watanabe}}, \citenamefont
  {{Taniguchi}}, \citenamefont {{Vignale}}, \citenamefont {{Polini}},
  \citenamefont {{Hone}}, \citenamefont {{Hillenbrand}},\ and\ \citenamefont
  {{Koppens}}}]{woessner2015}%
  \BibitemOpen
  \bibfield  {author} {\bibinfo {author} {\bibfnamefont {A.}~\bibnamefont
  {{Woessner}}}, \bibinfo {author} {\bibfnamefont {M.~B.}\ \bibnamefont
  {{Lundeberg}}}, \bibinfo {author} {\bibfnamefont {Y.}~\bibnamefont {{Gao}}},
  \bibinfo {author} {\bibfnamefont {A.}~\bibnamefont {{Principi}}}, \bibinfo
  {author} {\bibfnamefont {P.}~\bibnamefont {{Alonso-Gonz{\'a}lez}}}, \bibinfo
  {author} {\bibfnamefont {M.}~\bibnamefont {{Carrega}}}, \bibinfo {author}
  {\bibfnamefont {K.}~\bibnamefont {{Watanabe}}}, \bibinfo {author}
  {\bibfnamefont {T.}~\bibnamefont {{Taniguchi}}}, \bibinfo {author}
  {\bibfnamefont {G.}~\bibnamefont {{Vignale}}}, \bibinfo {author}
  {\bibfnamefont {M.}~\bibnamefont {{Polini}}}, \bibinfo {author}
  {\bibfnamefont {J.}~\bibnamefont {{Hone}}}, \bibinfo {author} {\bibfnamefont
  {R.}~\bibnamefont {{Hillenbrand}}}, \ and\ \bibinfo {author} {\bibfnamefont
  {F.~H.~L.}\ \bibnamefont {{Koppens}}},\ }\href {\doibase 10.1038/nmat4169}
  {\bibfield  {journal} {\bibinfo  {journal} {Nature Materials}\ }\textbf
  {\bibinfo {volume} {14}},\ \bibinfo {pages} {421} (\bibinfo {year}
  {2015})}\BibitemShut {NoStop}%
\bibitem [{\citenamefont {Dai}\ \emph {et~al.}(2018)\citenamefont {Dai},
  \citenamefont {Tymchenko}, \citenamefont {Xu}, \citenamefont {Tran},
  \citenamefont {Yang}, \citenamefont {Ma}, \citenamefont {Watanabe},
  \citenamefont {Taniguchi}, \citenamefont {Jarillo-Herrero}, \citenamefont
  {Aharonovich}, \citenamefont {Basov}, \citenamefont {Tao},\ and\
  \citenamefont {Alù}}]{Dai2018}%
  \BibitemOpen
  \bibfield  {author} {\bibinfo {author} {\bibfnamefont {S.}~\bibnamefont
  {Dai}}, \bibinfo {author} {\bibfnamefont {M.}~\bibnamefont {Tymchenko}},
  \bibinfo {author} {\bibfnamefont {Z.-Q.}\ \bibnamefont {Xu}}, \bibinfo
  {author} {\bibfnamefont {T.~T.}\ \bibnamefont {Tran}}, \bibinfo {author}
  {\bibfnamefont {Y.}~\bibnamefont {Yang}}, \bibinfo {author} {\bibfnamefont
  {Q.}~\bibnamefont {Ma}}, \bibinfo {author} {\bibfnamefont {K.}~\bibnamefont
  {Watanabe}}, \bibinfo {author} {\bibfnamefont {T.}~\bibnamefont {Taniguchi}},
  \bibinfo {author} {\bibfnamefont {P.}~\bibnamefont {Jarillo-Herrero}},
  \bibinfo {author} {\bibfnamefont {I.}~\bibnamefont {Aharonovich}}, \bibinfo
  {author} {\bibfnamefont {D.~N.}\ \bibnamefont {Basov}}, \bibinfo {author}
  {\bibfnamefont {T.~H.}\ \bibnamefont {Tao}}, \ and\ \bibinfo {author}
  {\bibfnamefont {A.}~\bibnamefont {Alù}},\ }\href {\doibase
  10.1021/acs.nanolett.8b02162} {\bibfield  {journal} {\bibinfo  {journal}
  {Nano Letters}\ }\textbf {\bibinfo {volume} {18}},\ \bibinfo {pages} {5205}
  (\bibinfo {year} {2018})}\BibitemShut {NoStop}%
\bibitem [{\citenamefont {Abbasirad}\ \emph {et~al.}(2019)\citenamefont
  {Abbasirad}, \citenamefont {Berzins}, \citenamefont {Kollin}, \citenamefont
  {Saravi}, \citenamefont {Janunts}, \citenamefont {Setzpfandt},\ and\
  \citenamefont {Pertsch}}]{Najmeh}%
  \BibitemOpen
  \bibfield  {author} {\bibinfo {author} {\bibfnamefont {N.}~\bibnamefont
  {Abbasirad}}, \bibinfo {author} {\bibfnamefont {J.}~\bibnamefont {Berzins}},
  \bibinfo {author} {\bibfnamefont {K.}~\bibnamefont {Kollin}}, \bibinfo
  {author} {\bibfnamefont {S.}~\bibnamefont {Saravi}}, \bibinfo {author}
  {\bibfnamefont {N.}~\bibnamefont {Janunts}}, \bibinfo {author} {\bibfnamefont
  {F.}~\bibnamefont {Setzpfandt}}, \ and\ \bibinfo {author} {\bibfnamefont
  {T.}~\bibnamefont {Pertsch}},\ }\href {\doibase 10.1063/1.5084946} {\bibfield
   {journal} {\bibinfo  {journal} {Review of Scientific Instruments}\ }\textbf
  {\bibinfo {volume} {90}},\ \bibinfo {pages} {053705} (\bibinfo {year}
  {2019})}\BibitemShut {NoStop}%
\bibitem [{\citenamefont {Frischwasser}\ \emph {et~al.}(2021)\citenamefont
  {Frischwasser}, \citenamefont {Cohen}, \citenamefont {Kher-Alden},
  \citenamefont {Dolev}, \citenamefont {Tsesses},\ and\ \citenamefont
  {Bartal}}]{frischwasser2021}%
  \BibitemOpen
  \bibfield  {author} {\bibinfo {author} {\bibfnamefont {K.}~\bibnamefont
  {Frischwasser}}, \bibinfo {author} {\bibfnamefont {K.}~\bibnamefont {Cohen}},
  \bibinfo {author} {\bibfnamefont {J.}~\bibnamefont {Kher-Alden}}, \bibinfo
  {author} {\bibfnamefont {S.}~\bibnamefont {Dolev}}, \bibinfo {author}
  {\bibfnamefont {S.}~\bibnamefont {Tsesses}}, \ and\ \bibinfo {author}
  {\bibfnamefont {G.}~\bibnamefont {Bartal}},\ }\href
  {https://doi.org/10.1038/s41566-021-00782-2} {\bibfield  {journal} {\bibinfo
  {journal} {Nature Photonics}\ }\textbf {\bibinfo {volume} {15}},\ \bibinfo
  {pages} {442} (\bibinfo {year} {2021})}\BibitemShut {NoStop}%
\bibitem [{\citenamefont {Okoth}\ \emph {et~al.}(2019)\citenamefont {Okoth},
  \citenamefont {Cavanna}, \citenamefont {Santiago-Cruz},\ and\ \citenamefont
  {Chekhova}}]{okoth1}%
  \BibitemOpen
  \bibfield  {author} {\bibinfo {author} {\bibfnamefont {C.}~\bibnamefont
  {Okoth}}, \bibinfo {author} {\bibfnamefont {A.}~\bibnamefont {Cavanna}},
  \bibinfo {author} {\bibfnamefont {T.}~\bibnamefont {Santiago-Cruz}}, \ and\
  \bibinfo {author} {\bibfnamefont {M.~V.}\ \bibnamefont {Chekhova}},\ }\href
  {\doibase 10.1103/PhysRevLett.123.263602} {\bibfield  {journal} {\bibinfo
  {journal} {Physical Review Letters}\ }\textbf {\bibinfo {volume} {123}},\
  \bibinfo {pages} {263602} (\bibinfo {year} {2019})}\BibitemShut {NoStop}%
\bibitem [{\citenamefont {Santiago-Cruz}\ \emph
  {et~al.}(2021{\natexlab{a}})\citenamefont {Santiago-Cruz}, \citenamefont
  {Sultanov}, \citenamefont {Zhang}, \citenamefont {Krivitsky},\ and\
  \citenamefont {Chekhova}}]{santiago1}%
  \BibitemOpen
  \bibfield  {author} {\bibinfo {author} {\bibfnamefont {T.}~\bibnamefont
  {Santiago-Cruz}}, \bibinfo {author} {\bibfnamefont {V.}~\bibnamefont
  {Sultanov}}, \bibinfo {author} {\bibfnamefont {H.}~\bibnamefont {Zhang}},
  \bibinfo {author} {\bibfnamefont {L.~A.}\ \bibnamefont {Krivitsky}}, \ and\
  \bibinfo {author} {\bibfnamefont {M.~V.}\ \bibnamefont {Chekhova}},\ }\href
  {\doibase 10.1364/OL.411176} {\bibfield  {journal} {\bibinfo  {journal}
  {Optics Letters}\ }\textbf {\bibinfo {volume} {46}},\ \bibinfo {pages} {653}
  (\bibinfo {year} {2021}{\natexlab{a}})}\BibitemShut {NoStop}%
\bibitem [{\citenamefont {Okoth}\ \emph {et~al.}(2020)\citenamefont {Okoth},
  \citenamefont {Kovlakov}, \citenamefont {B\"onsel}, \citenamefont {Cavanna},
  \citenamefont {Straupe}, \citenamefont {Kulik},\ and\ \citenamefont
  {Chekhova}}]{okoth2}%
  \BibitemOpen
  \bibfield  {author} {\bibinfo {author} {\bibfnamefont {C.}~\bibnamefont
  {Okoth}}, \bibinfo {author} {\bibfnamefont {E.}~\bibnamefont {Kovlakov}},
  \bibinfo {author} {\bibfnamefont {F.}~\bibnamefont {B\"onsel}}, \bibinfo
  {author} {\bibfnamefont {A.}~\bibnamefont {Cavanna}}, \bibinfo {author}
  {\bibfnamefont {S.}~\bibnamefont {Straupe}}, \bibinfo {author} {\bibfnamefont
  {S.~P.}\ \bibnamefont {Kulik}}, \ and\ \bibinfo {author} {\bibfnamefont
  {M.~V.}\ \bibnamefont {Chekhova}},\ }\href {\doibase
  10.1103/PhysRevA.101.011801} {\bibfield  {journal} {\bibinfo  {journal}
  {Physical review A}\ }\textbf {\bibinfo {volume} {101}},\ \bibinfo {pages}
  {011801(R)} (\bibinfo {year} {2020})}\BibitemShut {NoStop}%
\bibitem [{\citenamefont {Wang}\ \emph {et~al.}(2009)\citenamefont {Wang},
  \citenamefont {Rodr\'{\i}guez}, \citenamefont {Albers}, \citenamefont
  {Ahorinta}, \citenamefont {Sipe},\ and\ \citenamefont {Kauranen}}]{wang2009}%
  \BibitemOpen
  \bibfield  {author} {\bibinfo {author} {\bibfnamefont {F.~X.}\ \bibnamefont
  {Wang}}, \bibinfo {author} {\bibfnamefont {F.~J.}\ \bibnamefont
  {Rodr\'{\i}guez}}, \bibinfo {author} {\bibfnamefont {W.~M.}\ \bibnamefont
  {Albers}}, \bibinfo {author} {\bibfnamefont {R.}~\bibnamefont {Ahorinta}},
  \bibinfo {author} {\bibfnamefont {J.~E.}\ \bibnamefont {Sipe}}, \ and\
  \bibinfo {author} {\bibfnamefont {M.}~\bibnamefont {Kauranen}},\ }\href
  {\doibase 10.1103/PhysRevB.80.233402} {\bibfield  {journal} {\bibinfo
  {journal} {Physical Review B}\ }\textbf {\bibinfo {volume} {80}},\ \bibinfo
  {pages} {233402} (\bibinfo {year} {2009})}\BibitemShut {NoStop}%
\bibitem [{\citenamefont {Saleh}\ \emph {et~al.}(2018)\citenamefont {Saleh},
  \citenamefont {Vezzoli}, \citenamefont {Caspani}, \citenamefont {Branny},
  \citenamefont {Kumar}, \citenamefont {Gerardot},\ and\ \citenamefont
  {Faccio}}]{saleh}%
  \BibitemOpen
  \bibfield  {author} {\bibinfo {author} {\bibfnamefont {H.~D.}\ \bibnamefont
  {Saleh}}, \bibinfo {author} {\bibfnamefont {S.}~\bibnamefont {Vezzoli}},
  \bibinfo {author} {\bibfnamefont {L.}~\bibnamefont {Caspani}}, \bibinfo
  {author} {\bibfnamefont {A.}~\bibnamefont {Branny}}, \bibinfo {author}
  {\bibfnamefont {S.}~\bibnamefont {Kumar}}, \bibinfo {author} {\bibfnamefont
  {B.~D.}\ \bibnamefont {Gerardot}}, \ and\ \bibinfo {author} {\bibfnamefont
  {D.}~\bibnamefont {Faccio}},\ }\href
  {https://doi.org/10.1038/s41598-018-22270-4} {\bibfield  {journal} {\bibinfo
  {journal} {Scientific reports}\ }\textbf {\bibinfo {volume} {8}},\ \bibinfo
  {pages} {1} (\bibinfo {year} {2018})}\BibitemShut {NoStop}%
\bibitem [{\citenamefont {Santiago-Cruz}\ \emph
  {et~al.}(2021{\natexlab{b}})\citenamefont {Santiago-Cruz}, \citenamefont
  {Fedotova}, \citenamefont {Sultanov}, \citenamefont {Weissflog},
  \citenamefont {Arslan}, \citenamefont {Younesi}, \citenamefont {Pertsch},
  \citenamefont {Staude}, \citenamefont {Setzpfandt},\ and\ \citenamefont
  {Chekhova}}]{Anita}%
  \BibitemOpen
  \bibfield  {author} {\bibinfo {author} {\bibfnamefont {T.}~\bibnamefont
  {Santiago-Cruz}}, \bibinfo {author} {\bibfnamefont {A.}~\bibnamefont
  {Fedotova}}, \bibinfo {author} {\bibfnamefont {V.}~\bibnamefont {Sultanov}},
  \bibinfo {author} {\bibfnamefont {M.~A.}\ \bibnamefont {Weissflog}}, \bibinfo
  {author} {\bibfnamefont {D.}~\bibnamefont {Arslan}}, \bibinfo {author}
  {\bibfnamefont {M.}~\bibnamefont {Younesi}}, \bibinfo {author} {\bibfnamefont
  {T.}~\bibnamefont {Pertsch}}, \bibinfo {author} {\bibfnamefont
  {I.}~\bibnamefont {Staude}}, \bibinfo {author} {\bibfnamefont
  {F.}~\bibnamefont {Setzpfandt}}, \ and\ \bibinfo {author} {\bibfnamefont
  {M.}~\bibnamefont {Chekhova}},\ }\href
  {https://doi.org/10.1021/acs.nanolett.1c01125} {\bibfield  {journal}
  {\bibinfo  {journal} {Nano Letters}\ } (\bibinfo {year}
  {2021}{\natexlab{b}})}\BibitemShut {NoStop}%
\bibitem [{\citenamefont {Poddubny}\ \emph {et~al.}(2016)\citenamefont
  {Poddubny}, \citenamefont {Iorsh},\ and\ \citenamefont
  {Sukhorukov}}]{poddubny}%
  \BibitemOpen
  \bibfield  {author} {\bibinfo {author} {\bibfnamefont {A.~N.}\ \bibnamefont
  {Poddubny}}, \bibinfo {author} {\bibfnamefont {I.~V.}\ \bibnamefont {Iorsh}},
  \ and\ \bibinfo {author} {\bibfnamefont {A.~A.}\ \bibnamefont {Sukhorukov}},\
  }\href {https://link.aps.org/doi/10.1103/PhysRevLett.117.123901} {\bibfield
  {journal} {\bibinfo  {journal} {Physical Review Letters}\ }\textbf {\bibinfo
  {volume} {117}},\ \bibinfo {pages} {123901} (\bibinfo {year}
  {2016})}\BibitemShut {NoStop}%
\bibitem [{\citenamefont {Saravi}(2018)}]{SaraviThesis}%
  \BibitemOpen
  \bibfield  {author} {\bibinfo {author} {\bibfnamefont {S.}~\bibnamefont
  {Saravi}},\ }\emph {\bibinfo {title} {Photon-pair generation in photonic
  crystal waveguides}},\ \href {\doibase 10.22032/dbt.38074} {Ph.D. thesis},\
  \bibinfo  {school} {Friedrich-Schiller-Universit{\"a}t Jena} (\bibinfo {year}
  {2018})\BibitemShut {NoStop}%
\bibitem [{\citenamefont {Caz{\'e}}\ \emph {et~al.}(2013)\citenamefont
  {Caz{\'e}}, \citenamefont {Pierrat},\ and\ \citenamefont {Carminati}}]{caze}%
  \BibitemOpen
  \bibfield  {author} {\bibinfo {author} {\bibfnamefont {A.}~\bibnamefont
  {Caz{\'e}}}, \bibinfo {author} {\bibfnamefont {R.}~\bibnamefont {Pierrat}}, \
  and\ \bibinfo {author} {\bibfnamefont {R.}~\bibnamefont {Carminati}},\ }\href
  {https://link.aps.org/doi/10.1103/PhysRevLett.110.063903} {\bibfield
  {journal} {\bibinfo  {journal} {Physical Review Letters}\ }\textbf {\bibinfo
  {volume} {110}},\ \bibinfo {pages} {063903} (\bibinfo {year}
  {2013})}\BibitemShut {NoStop}%
\bibitem [{\citenamefont {Sauvan}\ \emph {et~al.}(2014)\citenamefont {Sauvan},
  \citenamefont {Hugonin}, \citenamefont {Carminati},\ and\ \citenamefont
  {Lalanne}}]{sauvan}%
  \BibitemOpen
  \bibfield  {author} {\bibinfo {author} {\bibfnamefont {C.}~\bibnamefont
  {Sauvan}}, \bibinfo {author} {\bibfnamefont {J.-P.}\ \bibnamefont {Hugonin}},
  \bibinfo {author} {\bibfnamefont {R.}~\bibnamefont {Carminati}}, \ and\
  \bibinfo {author} {\bibfnamefont {P.}~\bibnamefont {Lalanne}},\ }\href
  {https://link.aps.org/doi/10.1103/PhysRevA.89.043825} {\bibfield  {journal}
  {\bibinfo  {journal} {Physical Review A}\ }\textbf {\bibinfo {volume} {89}},\
  \bibinfo {pages} {043825} (\bibinfo {year} {2014})}\BibitemShut {NoStop}%
\bibitem [{\citenamefont {Carminati}\ \emph {et~al.}(2015)\citenamefont
  {Carminati}, \citenamefont {Caz{\'e}}, \citenamefont {Cao}, \citenamefont
  {Peragut}, \citenamefont {Krachmalnicoff}, \citenamefont {Pierrat},\ and\
  \citenamefont {De~Wilde}}]{carminati2015}%
  \BibitemOpen
  \bibfield  {author} {\bibinfo {author} {\bibfnamefont {R.}~\bibnamefont
  {Carminati}}, \bibinfo {author} {\bibfnamefont {A.}~\bibnamefont {Caz{\'e}}},
  \bibinfo {author} {\bibfnamefont {D.}~\bibnamefont {Cao}}, \bibinfo {author}
  {\bibfnamefont {F.}~\bibnamefont {Peragut}}, \bibinfo {author} {\bibfnamefont
  {V.}~\bibnamefont {Krachmalnicoff}}, \bibinfo {author} {\bibfnamefont
  {R.}~\bibnamefont {Pierrat}}, \ and\ \bibinfo {author} {\bibfnamefont
  {Y.}~\bibnamefont {De~Wilde}},\ }\href
  {https://www.sciencedirect.com/science/article/pii/S0167572914000338}
  {\bibfield  {journal} {\bibinfo  {journal} {Surface Science Reports}\
  }\textbf {\bibinfo {volume} {70}},\ \bibinfo {pages} {1} (\bibinfo {year}
  {2015})}\BibitemShut {NoStop}%
\bibitem [{\citenamefont {Kumar}\ \emph {et~al.}(2020)\citenamefont {Kumar},
  \citenamefont {Saravi}, \citenamefont {Pertsch},\ and\ \citenamefont
  {Setzpfandt}}]{pawan}%
  \BibitemOpen
  \bibfield  {author} {\bibinfo {author} {\bibfnamefont {P.}~\bibnamefont
  {Kumar}}, \bibinfo {author} {\bibfnamefont {S.}~\bibnamefont {Saravi}},
  \bibinfo {author} {\bibfnamefont {T.}~\bibnamefont {Pertsch}}, \ and\
  \bibinfo {author} {\bibfnamefont {F.}~\bibnamefont {Setzpfandt}},\ }\href
  {https://link.aps.org/doi/10.1103/PhysRevA.101.053860} {\bibfield  {journal}
  {\bibinfo  {journal} {Physical Review A}\ }\textbf {\bibinfo {volume}
  {101}},\ \bibinfo {pages} {053860} (\bibinfo {year} {2020})}\BibitemShut
  {NoStop}%
\bibitem [{\citenamefont {Castani\'{e}}\ \emph {et~al.}(2010)\citenamefont
  {Castani\'{e}}, \citenamefont {Boffety},\ and\ \citenamefont
  {Carminati}}]{Castanie}%
  \BibitemOpen
  \bibfield  {author} {\bibinfo {author} {\bibfnamefont {E.}~\bibnamefont
  {Castani\'{e}}}, \bibinfo {author} {\bibfnamefont {M.}~\bibnamefont
  {Boffety}}, \ and\ \bibinfo {author} {\bibfnamefont {R.}~\bibnamefont
  {Carminati}},\ }\href {\doibase 10.1364/OL.35.000291} {\bibfield  {journal}
  {\bibinfo  {journal} {Optics Letters}\ }\textbf {\bibinfo {volume} {35}},\
  \bibinfo {pages} {291} (\bibinfo {year} {2010})}\BibitemShut {NoStop}%
\bibitem [{\citenamefont {De~Vries}\ \emph {et~al.}(1998)\citenamefont
  {De~Vries}, \citenamefont {Van~Coevorden},\ and\ \citenamefont
  {Lagendijk}}]{vries}%
  \BibitemOpen
  \bibfield  {author} {\bibinfo {author} {\bibfnamefont {P.}~\bibnamefont
  {De~Vries}}, \bibinfo {author} {\bibfnamefont {D.~V.}\ \bibnamefont
  {Van~Coevorden}}, \ and\ \bibinfo {author} {\bibfnamefont {A.}~\bibnamefont
  {Lagendijk}},\ }\href {https://link.aps.org/doi/10.1103/RevModPhys.70.447}
  {\bibfield  {journal} {\bibinfo  {journal} {Reviews of modern physics}\
  }\textbf {\bibinfo {volume} {70}},\ \bibinfo {pages} {447} (\bibinfo {year}
  {1998})}\BibitemShut {NoStop}%
\bibitem [{\citenamefont {Carminati}\ \emph {et~al.}(2006)\citenamefont
  {Carminati}, \citenamefont {Greffet}, \citenamefont {Henkel},\ and\
  \citenamefont {Vigoureux}}]{carminati2006}%
  \BibitemOpen
  \bibfield  {author} {\bibinfo {author} {\bibfnamefont {R.}~\bibnamefont
  {Carminati}}, \bibinfo {author} {\bibfnamefont {J.-J.}\ \bibnamefont
  {Greffet}}, \bibinfo {author} {\bibfnamefont {C.}~\bibnamefont {Henkel}}, \
  and\ \bibinfo {author} {\bibfnamefont {J.-M.}\ \bibnamefont {Vigoureux}},\
  }\href {https://www.sciencedirect.com/science/article/pii/S0030401805013489}
  {\bibfield  {journal} {\bibinfo  {journal} {Optics Communications}\ }\textbf
  {\bibinfo {volume} {261}},\ \bibinfo {pages} {368} (\bibinfo {year}
  {2006})}\BibitemShut {NoStop}%
\bibitem [{\citenamefont {{Myers}}\ \emph {et~al.}(2018)\citenamefont
  {{Myers}}, \citenamefont {{Tonkyn}}, \citenamefont {{Danby}}, \citenamefont
  {{Taubman}}, \citenamefont {{Bernacki}}, \citenamefont {{Birnbaum}},
  \citenamefont {{Sharpe}},\ and\ \citenamefont
  {{Johnson}}}]{myers2018accurate}%
  \BibitemOpen
  \bibfield  {author} {\bibinfo {author} {\bibfnamefont {T.~L.}\ \bibnamefont
  {{Myers}}}, \bibinfo {author} {\bibfnamefont {R.~G.}\ \bibnamefont
  {{Tonkyn}}}, \bibinfo {author} {\bibfnamefont {T.~O.}\ \bibnamefont
  {{Danby}}}, \bibinfo {author} {\bibfnamefont {M.~S.}\ \bibnamefont
  {{Taubman}}}, \bibinfo {author} {\bibfnamefont {B.~E.}\ \bibnamefont
  {{Bernacki}}}, \bibinfo {author} {\bibfnamefont {J.~C.}\ \bibnamefont
  {{Birnbaum}}}, \bibinfo {author} {\bibfnamefont {S.~W.}\ \bibnamefont
  {{Sharpe}}}, \ and\ \bibinfo {author} {\bibfnamefont {T.~J.}\ \bibnamefont
  {{Johnson}}},\ }\href {\doibase 10.1177/0003702817742848} {\bibfield
  {journal} {\bibinfo  {journal} {Applied Spectroscopy}\ }\textbf {\bibinfo
  {volume} {72}},\ \bibinfo {pages} {535} (\bibinfo {year} {2018})}\BibitemShut
  {NoStop}%
\bibitem [{sup()}]{supplement}%
  \BibitemOpen
  \href@noop {} {\bibinfo  {journal} {See Supplemental Material at [URL] for
  analytical expressions of the different Green's functions involved in the
  numerical calculations, on the different contributions of detection
  polarization and dipolar orientation of the object to the total point-spread
  function, and the calculation of photon generation rates for the process}\
  }\BibitemShut {NoStop}%
\bibitem [{\citenamefont {des Francs}\ \emph {et~al.}(2008)\citenamefont {des
  Francs}, \citenamefont {Bouhelier}, \citenamefont {Finot}, \citenamefont
  {Weeber}, \citenamefont {Dereux}, \citenamefont {Girard},\ and\ \citenamefont
  {Dujardin}}]{Colasdes}%
  \BibitemOpen
\bibfield  {journal} {  }\bibfield  {author} {\bibinfo {author} {\bibfnamefont
  {G.~C.}\ \bibnamefont {des Francs}}, \bibinfo {author} {\bibfnamefont
  {A.}~\bibnamefont {Bouhelier}}, \bibinfo {author} {\bibfnamefont
  {E.}~\bibnamefont {Finot}}, \bibinfo {author} {\bibfnamefont {J.~C.}\
  \bibnamefont {Weeber}}, \bibinfo {author} {\bibfnamefont {A.}~\bibnamefont
  {Dereux}}, \bibinfo {author} {\bibfnamefont {C.}~\bibnamefont {Girard}}, \
  and\ \bibinfo {author} {\bibfnamefont {E.}~\bibnamefont {Dujardin}},\ }\href
  {\doibase 10.1364/OE.16.017654} {\bibfield  {journal} {\bibinfo  {journal}
  {Optics Express}\ }\textbf {\bibinfo {volume} {16}},\ \bibinfo {pages}
  {17654} (\bibinfo {year} {2008})}\BibitemShut {NoStop}%
\bibitem [{\citenamefont {Vandenbem}\ \emph {et~al.}(2010)\citenamefont
  {Vandenbem}, \citenamefont {Brayer}, \citenamefont {Froufe-P\'erez},\ and\
  \citenamefont {Carminati}}]{Vandenbem}%
  \BibitemOpen
  \bibfield  {author} {\bibinfo {author} {\bibfnamefont {C.}~\bibnamefont
  {Vandenbem}}, \bibinfo {author} {\bibfnamefont {D.}~\bibnamefont {Brayer}},
  \bibinfo {author} {\bibfnamefont {L.~S.}\ \bibnamefont {Froufe-P\'erez}}, \
  and\ \bibinfo {author} {\bibfnamefont {R.}~\bibnamefont {Carminati}},\ }\href
  {\doibase 10.1103/PhysRevB.81.085444} {\bibfield  {journal} {\bibinfo
  {journal} {Physical Review B}\ }\textbf {\bibinfo {volume} {81}},\ \bibinfo
  {pages} {085444} (\bibinfo {year} {2010})}\BibitemShut {NoStop}%
\bibitem [{\citenamefont {Vielma}\ and\ \citenamefont {Leung}(2007)}]{Vielma}%
  \BibitemOpen
  \bibfield  {author} {\bibinfo {author} {\bibfnamefont {J.}~\bibnamefont
  {Vielma}}\ and\ \bibinfo {author} {\bibfnamefont {P.~T.}\ \bibnamefont
  {Leung}},\ }\href {\doibase 10.1063/1.2734549} {\bibfield  {journal}
  {\bibinfo  {journal} {The Journal of Chemical Physics}\ }\textbf {\bibinfo
  {volume} {126}},\ \bibinfo {pages} {194704} (\bibinfo {year}
  {2007})}\BibitemShut {NoStop}%
\bibitem [{\citenamefont {Dickson}\ \emph {et~al.}(1998)\citenamefont
  {Dickson}, \citenamefont {Norris},\ and\ \citenamefont {Moerner}}]{Dickson}%
  \BibitemOpen
  \bibfield  {author} {\bibinfo {author} {\bibfnamefont {R.~M.}\ \bibnamefont
  {Dickson}}, \bibinfo {author} {\bibfnamefont {D.~J.}\ \bibnamefont {Norris}},
  \ and\ \bibinfo {author} {\bibfnamefont {W.~E.}\ \bibnamefont {Moerner}},\
  }\href {\doibase 10.1103/PhysRevLett.81.5322} {\bibfield  {journal} {\bibinfo
   {journal} {Physical Review Letters}\ }\textbf {\bibinfo {volume} {81}},\
  \bibinfo {pages} {5322} (\bibinfo {year} {1998})}\BibitemShut {NoStop}%
\bibitem [{\citenamefont {Lieb}\ \emph {et~al.}(2004)\citenamefont {Lieb},
  \citenamefont {Zavislan},\ and\ \citenamefont {Novotny}}]{Lieb}%
  \BibitemOpen
  \bibfield  {author} {\bibinfo {author} {\bibfnamefont {M.~A.}\ \bibnamefont
  {Lieb}}, \bibinfo {author} {\bibfnamefont {J.~M.}\ \bibnamefont {Zavislan}},
  \ and\ \bibinfo {author} {\bibfnamefont {L.}~\bibnamefont {Novotny}},\ }\href
  {\doibase 10.1364/JOSAB.21.001210} {\bibfield  {journal} {\bibinfo  {journal}
  {JOSA B}\ }\textbf {\bibinfo {volume} {21}},\ \bibinfo {pages} {1210}
  (\bibinfo {year} {2004})}\BibitemShut {NoStop}%
\bibitem [{\citenamefont {Born}\ and\ \citenamefont
  {Wolf}(2013)}]{born2013principles}%
  \BibitemOpen
  \bibfield  {author} {\bibinfo {author} {\bibfnamefont {M.}~\bibnamefont
  {Born}}\ and\ \bibinfo {author} {\bibfnamefont {E.}~\bibnamefont {Wolf}},\
  }\href {https://books.google.de/books?id=HY-GDAAAQBAJ} {\emph {\bibinfo
  {title} {Principles of Optics: Electromagnetic Theory of Propagation,
  Interference and Diffraction of Light}}}\ (\bibinfo  {publisher} {Elsevier
  Science},\ \bibinfo {year} {2013})\BibitemShut {NoStop}%
\bibitem [{\citenamefont {Schneeloch}\ and\ \citenamefont
  {Howell}(2016)}]{howell}%
  \BibitemOpen
  \bibfield  {author} {\bibinfo {author} {\bibfnamefont {J.}~\bibnamefont
  {Schneeloch}}\ and\ \bibinfo {author} {\bibfnamefont {J.~C.}\ \bibnamefont
  {Howell}},\ }\href {\doibase 10.1088/2040-8978/18/5/053501} {\bibfield
  {journal} {\bibinfo  {journal} {{IOP} Publishing}\ }\textbf {\bibinfo
  {volume} {18}},\ \bibinfo {pages} {053501} (\bibinfo {year}
  {2016})}\BibitemShut {NoStop}%
\bibitem [{\citenamefont {Saravi}\ \emph {et~al.}(2017)\citenamefont {Saravi},
  \citenamefont {Poddubny}, \citenamefont {Pertsch}, \citenamefont
  {Setzpfandt},\ and\ \citenamefont {Sukhorukov}}]{sina}%
  \BibitemOpen
  \bibfield  {author} {\bibinfo {author} {\bibfnamefont {S.}~\bibnamefont
  {Saravi}}, \bibinfo {author} {\bibfnamefont {A.~N.}\ \bibnamefont
  {Poddubny}}, \bibinfo {author} {\bibfnamefont {T.}~\bibnamefont {Pertsch}},
  \bibinfo {author} {\bibfnamefont {F.}~\bibnamefont {Setzpfandt}}, \ and\
  \bibinfo {author} {\bibfnamefont {A.~A.}\ \bibnamefont {Sukhorukov}},\ }\href
  {\doibase 10.1364/OL.42.004724} {\bibfield  {journal} {\bibinfo  {journal}
  {Optics Letters}\ }\textbf {\bibinfo {volume} {42}},\ \bibinfo {pages} {4724}
  (\bibinfo {year} {2017})}\BibitemShut {NoStop}%
\bibitem [{\citenamefont {Saleh}\ and\ \citenamefont
  {Teich}(2019)}]{saleh2019fundamentals}%
  \BibitemOpen
  \bibfield  {author} {\bibinfo {author} {\bibfnamefont {B.}~\bibnamefont
  {Saleh}}\ and\ \bibinfo {author} {\bibfnamefont {M.}~\bibnamefont {Teich}},\
  }\href {https://books.google.de/books?id=rcqKDwAAQBAJ} {\emph {\bibinfo
  {title} {Fundamentals of Photonics}}},\ Wiley Series in Pure and Applied
  Optics\ (\bibinfo  {publisher} {Wiley},\ \bibinfo {year} {2019})\BibitemShut
  {NoStop}%
\bibitem [{\citenamefont {Shoji}\ \emph {et~al.}(1997)\citenamefont {Shoji},
  \citenamefont {Kondo}, \citenamefont {Kitamoto}, \citenamefont {Shirane},\
  and\ \citenamefont {Ito}}]{shoji}%
  \BibitemOpen
  \bibfield  {author} {\bibinfo {author} {\bibfnamefont {I.}~\bibnamefont
  {Shoji}}, \bibinfo {author} {\bibfnamefont {T.}~\bibnamefont {Kondo}},
  \bibinfo {author} {\bibfnamefont {A.}~\bibnamefont {Kitamoto}}, \bibinfo
  {author} {\bibfnamefont {M.}~\bibnamefont {Shirane}}, \ and\ \bibinfo
  {author} {\bibfnamefont {R.}~\bibnamefont {Ito}},\ }\href {\doibase
  10.1364/JOSAB.14.002268} {\bibfield  {journal} {\bibinfo  {journal} {JOSA B}\
  }\textbf {\bibinfo {volume} {14}},\ \bibinfo {pages} {2268} (\bibinfo {year}
  {1997})}\BibitemShut {NoStop}%
\bibitem [{\citenamefont {Huber}\ \emph {et~al.}(2009)\citenamefont {Huber},
  \citenamefont {Ziegler}, \citenamefont {K{\"o}ck},\ and\ \citenamefont
  {Hillenbrand}}]{huber2009infrared}%
  \BibitemOpen
  \bibfield  {author} {\bibinfo {author} {\bibfnamefont {A.}~\bibnamefont
  {Huber}}, \bibinfo {author} {\bibfnamefont {A.}~\bibnamefont {Ziegler}},
  \bibinfo {author} {\bibfnamefont {T.}~\bibnamefont {K{\"o}ck}}, \ and\
  \bibinfo {author} {\bibfnamefont {R.}~\bibnamefont {Hillenbrand}},\ }\href
  {https://doi.org/10.1038/nnano.2008.399} {\bibfield  {journal} {\bibinfo
  {journal} {Nature nanotechnology}\ }\textbf {\bibinfo {volume} {4}},\
  \bibinfo {pages} {153} (\bibinfo {year} {2009})}\BibitemShut {NoStop}%
\bibitem [{\citenamefont {{Celebrano}}\ \emph {et~al.}(2011)\citenamefont
  {{Celebrano}}, \citenamefont {{Kukura}}, \citenamefont {{Renn}},\ and\
  \citenamefont {{Sandoghdar}}}]{Celebrano2011}%
  \BibitemOpen
  \bibfield  {author} {\bibinfo {author} {\bibfnamefont {M.}~\bibnamefont
  {{Celebrano}}}, \bibinfo {author} {\bibfnamefont {P.}~\bibnamefont
  {{Kukura}}}, \bibinfo {author} {\bibfnamefont {A.}~\bibnamefont {{Renn}}}, \
  and\ \bibinfo {author} {\bibfnamefont {V.}~\bibnamefont {{Sandoghdar}}},\
  }\href {\doibase 10.1038/nphoton.2010.290} {\bibfield  {journal} {\bibinfo
  {journal} {Nature Photonics}\ }\textbf {\bibinfo {volume} {5}},\ \bibinfo
  {pages} {95} (\bibinfo {year} {2011})}\BibitemShut {NoStop}%
\bibitem [{\citenamefont {Knoll}\ and\ \citenamefont
  {Keilmann}(1999)}]{knoll1999near}%
  \BibitemOpen
  \bibfield  {author} {\bibinfo {author} {\bibfnamefont {B.}~\bibnamefont
  {Knoll}}\ and\ \bibinfo {author} {\bibfnamefont {F.}~\bibnamefont
  {Keilmann}},\ }\href {https://doi.org/10.1038/20154} {\bibfield  {journal}
  {\bibinfo  {journal} {Nature}\ }\textbf {\bibinfo {volume} {399}},\ \bibinfo
  {pages} {134} (\bibinfo {year} {1999})}\BibitemShut {NoStop}%
\bibitem [{\citenamefont {Mayet}\ \emph {et~al.}(2008)\citenamefont {Mayet},
  \citenamefont {Dazzi}, \citenamefont {Prazeres}, \citenamefont {Allot},
  \citenamefont {Glotin},\ and\ \citenamefont {Ortega}}]{mayet2008sub}%
  \BibitemOpen
  \bibfield  {author} {\bibinfo {author} {\bibfnamefont {C.}~\bibnamefont
  {Mayet}}, \bibinfo {author} {\bibfnamefont {A.}~\bibnamefont {Dazzi}},
  \bibinfo {author} {\bibfnamefont {R.}~\bibnamefont {Prazeres}}, \bibinfo
  {author} {\bibfnamefont {F.}~\bibnamefont {Allot}}, \bibinfo {author}
  {\bibfnamefont {F.}~\bibnamefont {Glotin}}, \ and\ \bibinfo {author}
  {\bibfnamefont {J.}~\bibnamefont {Ortega}},\ }\href
  {http://www.osapublishing.org/ol/abstract.cfm?URI=ol-33-14-1611} {\bibfield
  {journal} {\bibinfo  {journal} {Optics Letters}\ }\textbf {\bibinfo {volume}
  {33}},\ \bibinfo {pages} {1611} (\bibinfo {year} {2008})}\BibitemShut
  {NoStop}%
\bibitem [{\citenamefont {Ballout}\ \emph {et~al.}(2011)\citenamefont
  {Ballout}, \citenamefont {Krassen}, \citenamefont {Kopf}, \citenamefont
  {Ataka}, \citenamefont {Br{\"u}ndermann}, \citenamefont {Heberle},\ and\
  \citenamefont {Havenith}}]{ballout2011scanning}%
  \BibitemOpen
  \bibfield  {author} {\bibinfo {author} {\bibfnamefont {F.}~\bibnamefont
  {Ballout}}, \bibinfo {author} {\bibfnamefont {H.}~\bibnamefont {Krassen}},
  \bibinfo {author} {\bibfnamefont {I.}~\bibnamefont {Kopf}}, \bibinfo {author}
  {\bibfnamefont {K.}~\bibnamefont {Ataka}}, \bibinfo {author} {\bibfnamefont
  {E.}~\bibnamefont {Br{\"u}ndermann}}, \bibinfo {author} {\bibfnamefont
  {J.}~\bibnamefont {Heberle}}, \ and\ \bibinfo {author} {\bibfnamefont
  {M.}~\bibnamefont {Havenith}},\ }\href {https://doi.org/10.1021/nl301159v}
  {\bibfield  {journal} {\bibinfo  {journal} {Physical Chemistry Chemical
  Physics}\ }\textbf {\bibinfo {volume} {13}},\ \bibinfo {pages} {21432}
  (\bibinfo {year} {2011})}\BibitemShut {NoStop}%
\bibitem [{\citenamefont {{Huth}}\ \emph {et~al.}(2012)\citenamefont {{Huth}},
  \citenamefont {{Govyadinov}}, \citenamefont {{Amarie}}, \citenamefont
  {{Nuansing}}, \citenamefont {{Keilmann}},\ and\ \citenamefont
  {{Hillenbrand}}}]{huth2012nano}%
  \BibitemOpen
  \bibfield  {author} {\bibinfo {author} {\bibfnamefont {F.}~\bibnamefont
  {{Huth}}}, \bibinfo {author} {\bibfnamefont {A.}~\bibnamefont
  {{Govyadinov}}}, \bibinfo {author} {\bibfnamefont {S.}~\bibnamefont
  {{Amarie}}}, \bibinfo {author} {\bibfnamefont {W.}~\bibnamefont
  {{Nuansing}}}, \bibinfo {author} {\bibfnamefont {F.}~\bibnamefont
  {{Keilmann}}}, \ and\ \bibinfo {author} {\bibfnamefont {R.}~\bibnamefont
  {{Hillenbrand}}},\ }\href {\doibase 10.1021/nl301159v} {\bibfield  {journal}
  {\bibinfo  {journal} {Nano Letters}\ }\textbf {\bibinfo {volume} {12}},\
  \bibinfo {pages} {3973} (\bibinfo {year} {2012})}\BibitemShut {NoStop}%
\bibitem [{\citenamefont {{Elshaari}}\ \emph {et~al.}(2020)\citenamefont
  {{Elshaari}}, \citenamefont {{Pernice}}, \citenamefont {{Srinivasan}},
  \citenamefont {{Benson}},\ and\ \citenamefont {{Zwiller}}}]{Elshaari2020}%
  \BibitemOpen
  \bibfield  {author} {\bibinfo {author} {\bibfnamefont {A.~W.}\ \bibnamefont
  {{Elshaari}}}, \bibinfo {author} {\bibfnamefont {W.}~\bibnamefont
  {{Pernice}}}, \bibinfo {author} {\bibfnamefont {K.}~\bibnamefont
  {{Srinivasan}}}, \bibinfo {author} {\bibfnamefont {O.}~\bibnamefont
  {{Benson}}}, \ and\ \bibinfo {author} {\bibfnamefont {V.}~\bibnamefont
  {{Zwiller}}},\ }\href {\doibase 10.1038/s41566-020-0609-x} {\bibfield
  {journal} {\bibinfo  {journal} {Nature Photonics}\ }\textbf {\bibinfo
  {volume} {14}},\ \bibinfo {pages} {285} (\bibinfo {year} {2020})}\BibitemShut
  {NoStop}%
\end{thebibliography}


\end{document}